\documentclass[11pt]{article}
\usepackage{amsfonts}
\usepackage{amsmath}
\usepackage{amssymb}
\usepackage{geometry}
\usepackage{setspace}
\usepackage{float}
\usepackage{portland}
\usepackage{apacite}
\usepackage{apalike}
\usepackage{scalefnt}
\usepackage{sectsty}
\usepackage[USenglish]{babel}

\allsectionsfont{\normalsize}
\input{tcilatex}
\begin{document}

\title{%
\begingroup%
\scalefont{.7}%
\textbf{A}\ \textbf{Dirichlet Process Functional Approach to}\\
\textbf{Heteroscedastic-Consistent Covariance Estimation}\\
\textit{International Journal of Approximate Reasoning }Special Issue on
Bayesian Nonparametrics%
\endgroup%
}
\author{%
\begingroup%
\scalefont{.9}%
George Karabatsos%
\endgroup
\\
\begingroup%
\scalefont{.9}%
\textit{University of Illinois-Chicago}%
\endgroup%
}
\date{June 11, 2016}
\maketitle

\begin{abstract}
\begingroup%
\scalefont{1.01}%
The mixture of Dirichlet process (MDP)\ defines a flexible prior
distribution on the space of probability measures. This study shows that
ordinary least-squares (OLS) estimator,\ as a functional of the MDP
posterior distribution, has posterior mean given by weighted least-squares
(WLS), and has posterior covariance matrix given by the (weighted)\
heteroscedastic-consistent sandwich estimator. This is according to a pairs
bootstrap distribution approximation of the posterior, using a P\'{o}lya urn
scheme. Also, when the MDP\ prior baseline distribution is specified as a
product of independent probability measures, this WLS solution provides a
new type of generalized ridge regression estimator. Such an estimator can
handle multicollinear or singular design matrices even when the number of
covariates exceeds the sample size, and can shrinks the coefficient
estimates of irrelevant covariates towards zero, which makes it useful for
nonlinear regressions via basis expansions. Also, this MDP/OLS\ functional
methodology can be extended to methods for analyzing the sensitivity of the
heteroscedasticity-consistent causal effect size over a range of hidden
biases due to missing covariates omitted from the regression, and more
generally extended to a Vibration of Effects analysis. The methodology is
illustrated through the analysis of simulated and real data sets. Overall,
this study establishes new connections between Dirichlet process functional
inference, the bootstrap, consistent sandwich covariance estimation, ridge
shrinkage regression, WLS, and sensitivity analysis, to provide regression
methodology useful for inferences of the mean dependent response.\newline
\endgroup%
\end{abstract}

\section{Introduction}

When the linear regression model is misspecified due to the presence of
heteroscedasticity, the sampling covariance matrix of the ordinary
least-squares (OLS) estimator of the regression coefficients becomes
inconsistent. White's (1980\nocite{White80b}) sandwich covariance matrix
estimator is consistent even under heteroscedasticity, and does not require
any modeling specification for the form of the heteroscedasticity (see also
Eicker, 1963\nocite{Eicker63}, 1967\nocite{Eicker67}; Huber, 1967\nocite%
{Huber67}). Hence, the sandwich estimator is often referred to as
heteroscedastic-consistent or -robust.

White's article has profoundly impacted applied statistics and econometrics.
By June 2006, it was most cited by others in the peer-reviewed economics
literature since 1970 (Kim et al., 2006\nocite{KimMorseZingales06}), and
cited over 21,700 times according to a May 2016 internet search. However,
from a frequentist perspective, the sandwich estimator can exhibit downward
bias for small sample size ($n$) data sets containing observations with high
leverage\ on the OLS\ estimand (Chesher \&\ Jewitt, 1987\nocite%
{ChesherJewitt87}). This has led researchers to propose various
leverage-adjusted sandwich estimators (see MacKinnon, 2013\nocite%
{MacKinnon13}).

Recent studies have proposed Bayesian linear modeling methods that make use
of the sandwich estimator. M\"{u}ller (2013\nocite{Muller13}) showed that
regression coefficient inference has lower asymptotic frequentist risk when
using an artificial multivariate normal\ posterior distribution centered on
the maximum likelihood estimate with sandwich covariance matrix, compared to
the posterior distribution under the homoscedasticity assumption. Hoff and
Wakefield (2013\nocite{HoffWakefield13}) and Startz (2015\nocite{Startz15})
extended this approach by incorporating informative prior distributions, and
showed that the heteroscedastic-robust posterior can exhibit more
uncertainty than the posterior under the homoscedasticity assumption. Norets
(2015\nocite{Norets15}) proposed flexible Bayesian nonparametric (e.g.,
Gaussian process) models for the regression error terms, with the motivation
that fully Bayesian nonparametric, Dependent Dirichlet process (DDP)\
infinite-mixture regression models (e.g., DeIorio et al. 2004\nocite%
{DeIorioMullerRosnerMacEachern04};\ Dunson \&\ Park, 2008\nocite%
{DunsonPark08}) require a lot of data for reliable estimation results, and
require prior specification that is non-trivial in practice. Generally
speaking, each of these prior-informed Bayesian regression models requires
use of an Markov chain Monte Carlo (MCMC) sampling algorithm for estimating
the posterior distribution of the regression coefficients. MCMC\ can be
computationally-intensive for a large data set. Further, it may be argued
that if in a practical setting the primary aim is to perform
heteroscedastic-consistent inferences of linear regression coefficients,
then there is no point in using intensive Monte Carlo estimation methods
because OLS and sandwich estimators can already be numerically evaluated
(MacKinnon, 2013\nocite{MacKinnon13}).

Lancaster (2003\nocite{Lancaster03}) showed that the OLS estimator, as a
functional of the classical bootstrap (CB)\ (Efron, 1979\nocite{Efron79})
distribution, or of the Bayesian Bootstrap (BB)\ (Rubin, 1981\nocite{Rubin81}%
) distribution, has covariance matrix that is order $n^{-1}$ equivalent to
the sandwich estimator. Here, we refer to the \textit{pairs} bootstrap,
where each of the $n$ observations consists of the dependent variable
observation paired with its corresponding observations on $p$ covariates. In
the CB, the $n$ observations are assigned (single-trial)\ multinomial
(re)sampling probabilities (weights)\ of $1/n$ (resp.). In the BB, these $n$
sampling probabilities have a Dirichlet posterior distribution with
concentration parameters 1 (resp.), under an improper non-informative prior.
The pairs BB eas studied by Chamberlain and Imbens (2003\nocite%
{ChamberlainImbens03}), Szpiro et al. (2010\nocite{SzpiroRiceLumley10}), and
Taddy et al. (2015\nocite{TaddyEtAl15}).

Poirier (2011\nocite{Poirier11}) proposed a pairs BB\ approach that employs
an informative Dirichlet prior distribution that can be chosen to assign
positive support to data observations, and to imaginary observations. He
showed that the prior-informed pairs BB distribution of the OLS estimator
has posterior mean given by weighted least squares (WLS), and posterior
covariance matrix given by a weighted sandwich estimator, according to a
Taylor series\ approximation. He also showed that nearly all of the
frequentist-based leverage-adjusted sandwich estimators (mentioned earlier)\
can be characterized as assuming a particular Dirichlet prior distribution
that does not support imaginary observations, and can give rise to a
posterior distribution that places to much support to extreme sampling
probability weights. However, he observed that it is not necessarily easy in
practice to elicit an informative Dirichlet prior that supports imaginary
observations, and then concluded that more attractive informative prior
specifications await further research.

The Dirichlet process (DP)\ defines a flexible prior distribution on the
space of random probability measures (r.p.m.s)\ (distribution functions),
and is parameterized by a baseline distribution and a precision parameter ($%
\alpha $) which respectively control the mean and variance of the r.p.m.
(Ferguson, 1973\nocite{Ferguson73}). The DP prior has the conjugacy
property, in the sense that a data update of this prior leads to a posterior
distribution for the r.p.m. that is also a DP; and the DP\ is the only
process that has this conjugacy property in the class of homogeneous
normalized random measures with independent increments (James, et al., 2006%
\nocite{JamesLijoiPruenster06}). Also, the BB's Dirichlet posterior
distribution is the DP posterior distribution under a non-informative DP
prior with limiting zero precision parameter.

Cifarelli and Regazzini (1979\nocite{CifarelliRegazzini79a}) initiated a
line of research that deals with the problem of determining the expression
for the distribution of functionals of the Dirichlet process, with any
prescribed error of approximation (for reviews, see Regazzini et al., 2002%
\nocite{RegazziniGuglielmiDiNunno02}; Lijoi \& Prunster, 2009\nocite%
{LijoiPrunster09}). This research has primarily focused the mean and other
linear functionals of the DP.

This article studies approximations of the distribution of the OLS estimator
as a functional of the mixture of Dirichlet process (MDP) posterior
distribution. The MDP\ prior is a DP\ prior, with a hyperprior distribution
(at least)\ on the precision parameter (Antoniak, 1974\nocite{Antoniak74}).
Conditionally on the precision and baseline parameters, the DP posterior
distribution can be well-approximated by a bootstrap distribution that is
defined by the P\'{o}lya urn scheme characterization of the DP posterior
(Blackwell \&\ MacQueen, 1973\nocite{BlackwellMacQueen73}). Specifically, if
each resampled data set of the bootstrap procedure has sample size $n+\alpha
+1$, then the DP posterior and bootstrap distributions have identical means
and variances (Hjort, 1985\nocite{Hjort85}). This equality can be directly
verified analytically, thanks to the DP conjugacy property that allows for
explicit expressions of the posterior DP\ mean and variance. As a
consequence, for any well-behaved functional, including the OLS estimator,
the bootstrap distribution of the functional (via the P\'{o}lya urn scheme)
well-approximates the DP\ posterior distribution of the functional (Hjort,
1985\nocite{Hjort85}). By extension, this is true for the MDP\ posterior
distribution of the functional, after marginalizing out the posterior
distribution of the precision parameter. In this study we focus on the DP\
(MDP) because it is the only Gibbs-type prior that enables posterior
consistency for either continuous or discrete r.p.m.s (De Blasi et al., 2015%
\nocite{DeBlasiEtAl15}), while regression applications often involve the use
of continuous variables.

Ferguson (1973\nocite{Ferguson73}, p. 209)\ introduced the DP\ prior with
the motivation that the DP\ posterior distribution "should be manageable
analytically," and that the "support of the prior distribution should be
large-with respect to some suitable topology on the space of probability
distributions on the sample space." He then provided explicit analytical
solutions to a list of nonparametric statistical problems based on the DP
posterior, including the estimation of a distribution function, median,
quantiles, variance, covariance, and the probability that one variable
exceeds another.

The current article adds to his list by showing that the OLS estimator, as a
functional of the MDP\ posterior distribution, has posterior mean given by
WLS, and posterior covariance matrix given by a weighted
heteroscedastic-consistent sandwich estimator. This is according to a Taylor
series\ approximation of the pairs bootstrap (MDP\ posterior predictive)\
distribution, using the multivariate delta method.\ Under a non-informative
DP\ prior, this sandwich estimator closely approximates White's (1980\nocite%
{White80b}) original (unweighted) sandwich estimator. Also, it is shown that
if the MDP\ prior baseline distribution is specified as a product of
independent probability measures, then this WLS solution is the Bayesian
generalized ridge regression estimator (Hoerl \&\ Kennard, 1970\nocite%
{HoerlKennard70}). It is known that such an estimator can handle
multicollinear or singular covariate design matrices, even when the number
of covariates exceeds the sample size (i.e., $p>n$), while shrinking the
coefficient estimates of irrelevant covariates towards zero. These features
of ridge regression are useful for fitting nonlinear regressions via basis
expansions, and further ridge regression is tough to beat in terms of
predictive power (Griffin \&\ Brown, 2013\nocite{GriffinBrown13}). Clearly,
these posterior quantities (WLS\ and sandwich estimators) are analytically
manageable and permit fast computations even for large data sets. The
current study is the first to draw connections between the DP\ and Bayesian
ridge regression, and to provide heteroscedastic-consistent covariance
estimation for ridge regression.

The following sections elaborate on the main findings of this article.
Section 2 describes the specific MDP model that is employed, and presents
the model's key conditional and marginal posterior distributions. This
includes the posterior of the precision parameter as give by Nandram and\
Choi (2004\nocite{NandramChoi04}). Section 3 briefly reviews the key
properties and assumptions of the OLS and sandwich estimators. It then
establishes connections between the OLS estimator, ridge regression, and the
posterior moments of the MDP\ model, using imputation methods for imaginary
data that has the chosen baseline distribution for the MDP. Section 4
provides details about how the OLS functional of the MDP\ posterior is
approximated by the bootstrap distribution of this functional.

In practice, if the OLS\ (WLS)\ estimate of the regression coefficients is
inconsistent or biased, then the heteroscedastic-consistency of the sandwich
covariance estimator can become meaningless (Freedman, 2006\nocite%
{Freedman06}). This inconsistency results from correlation between
covariates and regression errors, implying a violation of the exogeneity
assumption of regression (Greene, 2012\nocite{Greene12}) and the presence of
hidden bias due to missing covariates ("confounders")\ omitted from the
regression equation (Rosenbaum, 2002\nocite{Rosenbaum02a}). Section 5
describes how the MDP/OLS\ functional methodology can easily incorporate
methods of sensitivity analysis (van der\ Weele \&\ Arah, 2011\nocite%
{VanderweeleArah11}), which aim to evaluate how much the causal effect size,
of a covariate of interest, varies over a hypothesized\ range of hidden
biases. In the current setting, the effect size is defined by the ratio of
the slope coefficient estimate of the covariate, over its
heteroscedastic-consistent posterior standard deviation.

Generally speaking, the MDP/OLS\ methodology can incorporate a vibration of
effects (VoE) analysis (Ioannidis, 2008\nocite{Ioannidis08}) in order to
assess how much the effect size differs (or vibrates) over different ways
that the analysis can be performed, for e.g., with respect to different:\
variables that are included and excluded in the analysis (statistical
adjustments); models used; definitions of outcomes and predictors; and
inclusion and exclusion criteria for the study population. VoE analysis
addresses the fact that an effect size estimator can display noticeable
variance (vibration) over different ways that the data analysis is done.
This variance can lead to bias if only a few chosen analyses are reported,
especially if the investigators have a preference for a particular result or
are influenced by optimism bias (Ioannidis et al. 2014\nocite%
{IoannidisEtAl14}, p.168). A recent study proposed a VoE analysis method for
regression settings (Patel, et al. 2015\nocite{PatelBurfordIoannidis15}),
which entails studying the variance of the effect size over all different
subsets of $K-1$ other (adjustment)\ covariates that may be included in the
regression. But as noted, this full enumeration approach is infeasible for
sufficiently large $K$. In this study consider a VoE analysis approach that
employs the Least Angle Regression (LARS) algorithm (Efron, et al. 2004%
\nocite{EfronHastieJohnstoneTibshirani04}). LARS provides a fast and
directed selection of covariates, yielding a path of $K+1$ regression
solutions that include $k=0,\ldots ,K$ covariates in the regression equation
(resp.), at the computational cost of a single OLS fit.

Section 6 describes a simulation study that evaluates the MDP/OLS\
functional methodology in terms of coverage rates of 95\%\ posterior
intervals of linear regression coefficients. These rates are studied over a
range of conditions of sample size, covariate distribution, degree of
heteroscedasticity, and choice of prior distribution for the MDP\ precision
parameter. Section 7 illustrates the functional methodology on two real data
sets. Section 8 concludes with some suggestions for future research,
including extensions of the methodology to other Bayesian nonparametric
priors.

\section{Mixture of Dirichlet Process Model}

Let $\mathbf{Z}_{n}=\mathbf{(\mathbf{z}}_{i}^{\intercal }=(\mathbf{x}%
_{i}^{\intercal },y_{i}\mathbf{))}_{i=1}^{n}=(\mathbf{X,y})=\mathbf{(X}_{n}%
\mathbf{,y}_{n}\mathbf{)}$ denote a data set (matrix)\ of $n$ observations
of the variable $\boldsymbol{Z}=(\boldsymbol{X},Y)$, including a dependent
variable $Y$ and $K$ covariates $\mathbf{x}=(x_{1},\ldots ,x_{k},\ldots
,x_{K})^{\intercal }$. The data set has $c_{n}\leq n$ distinct values
(clusters) $\mathbf{Z}_{c_{n}}^{\ast }=(\mathbf{X}_{c_{n}}^{\ast },\mathbf{y}%
_{c_{n}}^{\ast })=(\mathbf{z}_{c}^{\ast \intercal }=(\mathbf{x}_{c}^{\ast
\intercal },y_{c}^{\ast }))_{c=1}^{c_{n}\leq n},$ with frequency counts $%
\mathbf{n}_{c_{n}}=(n_{1},\ldots ,n_{c_{n}})^{\intercal }$ (resp.), and $%
\tsum\nolimits_{c=1}^{c_{n}}n_{c}=n$. Such a data set is assumed to consist
of $n$ exchangeable samples from an unknown distribution function $F$,
having space $\mathcal{F}_{\mathcal{Z}}=\{F\}$, the set of all probability
measures on $\mathcal{Z}=\{\boldsymbol{Z}\}\subset 
\mathbb{R}
^{K+1}$, according to the MDP model: 
\begin{subequations}
\label{MDP}
\begin{eqnarray}
\mathbf{z}_{i}\,|\,F &\sim &F,\text{ }i=1,\ldots ,n, \\
F\,|\,\alpha &\sim &\mathcal{DP}(\alpha ,F_{0}), \\
F_{0}(\mathbf{z}) &=&\mathrm{N}_{K+1}(\mathbf{x}_{i}^{\intercal },y_{i}\,|\,%
\mathbf{m}_{\mathbf{z}},\mathbf{V}_{\mathbf{z}})  \label{baseline} \\
\alpha &\sim &\pi (\alpha )\text{, }(\alpha >0)\text{.}  \label{MDPalpha}
\end{eqnarray}%
$\mathcal{DP}(\alpha ,F_{0})$ denotes the Dirichlet process (DP)\ prior
distribution on $\mathcal{F}_{\mathcal{Z}}$, with precision parameter $%
\alpha $, and baseline distribution $F_{0}$, specified as a $(K+1)$-variate
normal distribution with mean vector parameter $\mathbf{m}_{\mathbf{z}%
}=(m_{k})_{k=1}^{K+1}=(\mathbf{m}_{\mathbf{x}}^{\intercal
},m_{Y})^{\intercal }$ and covariance matrix parameter $\mathbf{V}_{\mathbf{z%
}}$.

In the current study, we focus on the \textit{ridge baseline prior}, defined
by: 
\end{subequations}
\begin{equation}
F_{0}(\mathbf{z})=\mathrm{N}(x_{1}\,|\,0,0)\tprod\nolimits_{k=2}^{K}\mathrm{N%
}(x_{k}\,|\,0,v_{\mathbf{x}k})\mathrm{N}(y\,|\,0,0),  \label{RidgeBase}
\end{equation}%
implying $\mathbf{m}_{\mathbf{z}}=(\mathbf{m}_{\mathbf{x}}^{\intercal
},m_{Y})^{\intercal }=\mathbf{0}_{K+1}^{\intercal }$, $\mathbf{V}_{\mathbf{z}%
}=\mathrm{diag}(0,v_{\mathbf{x}2},\ldots ,v_{\mathbf{x}K},0)$, with $\mathbf{%
0}_{K}$ a column vector of $K$ zeros. The \textit{unit ridge baseline prior}
further assumes $\mathbf{V}_{\mathbf{z}}=\mathrm{diag}(0,\mathbf{1}%
_{K-1}^{\intercal },0)$, with $\mathbf{1}_{K}$ a column vector of $K$ ones.
This study finds that each of these baseline distributions, along with $%
\alpha $, has connections with ridge regression, for reasons given in the
next section.

The precision parameter $\alpha $ in (\ref{MDPalpha}) of the MDP model (\ref%
{MDP}) represents a (prior) sample size for the number of imaginary
observations of $\mathbf{z}$, and is assigned a prior distribution with
p.d.f. $\pi (\alpha )$. In this study we consider the uniform prior p.d.f. $%
\mathrm{un}(\alpha \,|\,0,\xi )=\mathbf{1}(0<\alpha <\xi )/\xi $, where $%
\mathbf{1}(\cdot )$ is the indicator function; as well as a $\xi $-truncated
version of a Cauchy-type shrinkage prior p.d.f. $\pi (\alpha )=\mathbf{1}%
(0<\alpha <\xi )/(\alpha +1)^{2}$, $\alpha >0$ (Nandram \&\ Yin, 2016\nocite%
{NandramYin16}).

The conditional DP\ prior distribution $F\,|\,\alpha \sim \mathcal{DP}%
(\alpha ,F_{0})$ has a Dirichlet ($\mathrm{Di}$) distribution: 
\begin{equation}
F(B_{1}),\ldots ,F(B_{k})\,|\,\alpha \sim \mathrm{Di}_{k}(\alpha
F_{0}(B_{1}),\ldots ,\alpha F_{0}(B_{k})),
\end{equation}%
for all $k\geq 1$ partitions $B_{1},\ldots ,B_{k}$ of $\mathcal{Z}$, with
prior mean $\mathbb{E}[F(B)\,|\,\alpha ]=F_{0}(B)$ and variance $\mathbb{V}%
[F(B)\,|\,\alpha ]=F_{0}(B)\{1-F_{0}(B)\}/(\alpha +1)$ for $\forall B\in 
\mathcal{B}(\mathcal{Z})$ (Ferguson, 1973\nocite{Ferguson73}). The
probability (likelihood)\ distribution for the number of clusters is given
by (Antoniak, 1974\nocite{Antoniak74}):%
\begin{equation}
P(C_{n}=k\,|\,\alpha )=\dfrac{s_{n}(k)\alpha ^{k}\Gamma (\alpha )}{\Gamma
(\alpha +n)},\text{ }(\alpha >0),  \label{DP Cluster Like}
\end{equation}%
where the $s_{n}(k)$ are the signless Stirling numbers of the first kind
(Abramowitz \& Stegun, 1965\nocite{AbramowitzStegun65}).

The conditional posterior distribution of $F$ is also a DP, with $F\,|\,%
\mathbf{Z}_{n},\alpha \sim \mathcal{DP}(\alpha ,F_{0})$, and has Dirichlet
distribution: 
\begin{equation}
F(B_{1}),\ldots ,F(B_{k})\,|\,\mathbf{Z}_{n},\alpha \sim \mathrm{Di}%
_{k}(\alpha F_{0}(B_{1})+n\widehat{F}_{n}(B_{1}),\ldots ,\alpha
F_{0}(B_{k})+n\widehat{F}_{n}(B_{k})),  \label{DPpostCond1}
\end{equation}%
for all $k\geq 1$ partitions $B_{1},\ldots ,B_{k}$ of $\mathcal{Z}$, with
baseline distribution $F_{0}$ (\ref{baseline}); $\widehat{F}_{n}(\cdot
)=\tsum\nolimits_{c=1}^{c_{n}}\tfrac{n_{c}}{n}\delta _{\mathbf{z}_{c}^{\ast
}}(\cdot )$ is the empirical distribution function (e.d.f.)\ of the data, $%
\mathbf{Z}_{n}$; and $\delta _{z}(\cdot )$ is the degenerate probability
measure $\delta _{\mathbf{z}}(\mathbf{z})=1$, with $\delta _{\mathbf{z}}(B)=1
$ if $\mathbf{z}\in B$ (Ferguson, 1973\nocite{Ferguson73}). This posterior
distribution has conditional expectation and variance (resp.): 
\begin{subequations}
\begin{eqnarray}
\mathbb{E}[F(B)\,|\,\mathbf{Z}_{n},\alpha ] &=&\Pr (\mathbf{z}_{n+1}\in
B\,|\,\mathbf{Z}_{n},\alpha )  \label{EFgivAlphaBase} \\
&=&\tfrac{n}{\alpha +n}\widehat{F}_{n}(B)+\tfrac{\alpha }{\alpha +n}%
F_{0}(B):=\overline{F}_{\alpha }(B),  \label{EFgivAlphaBase2} \\
&=&\tsum\nolimits_{c=1}^{c_{n}}\tfrac{n_{c}}{\alpha +n}\delta _{\mathbf{z}%
_{c}^{\ast }}(B)+\tfrac{\alpha }{\alpha +n}F_{0}(B);  \label{EFgivAlphaBase3}
\\
\mathbb{V}[F(B)\,|\,\mathbf{Z}_{n},\alpha ] &=&\dfrac{\overline{F}_{\alpha
}(B)\{1-\overline{F}_{\alpha }(B)\}}{\alpha +n+1},\text{ \ }\forall B\in 
\mathcal{B}(\mathcal{Z}).  \label{VFgivAlphaBase}
\end{eqnarray}%
The posterior expectation (\ref{EFgivAlphaBase})-(\ref{EFgivAlphaBase3})
gives the posterior predictive probability of a new observation, $\mathbf{z}%
_{n+1}=(\mathbf{x}_{n+1}^{\intercal },y_{n+1})^{\intercal }\in \mathcal{Z}$,
according to the P\'{o}lya urn scheme (Blackwell \&\ MacQueen, 1973\nocite%
{BlackwellMacQueen73}). This scheme states that with probability $\tfrac{%
n_{c}}{\alpha +n},$ a new observation $\mathbf{z}_{n+1}=(\mathbf{x}%
_{n+1},y_{n+1})$ takes on value $\mathbf{z}_{c}^{\ast }=(\mathbf{x}%
_{c}^{\ast },y_{c}^{\ast })$ of an existing cluster $c$, for $c=1,\ldots
,c_{n}$; and otherwise with probability $\tfrac{\alpha }{\alpha +n}$, the
new observation $\mathbf{z}_{n+1}=(\mathbf{x}_{n+1},y_{n+1})$ is a sample $%
\mathbf{z}_{c_{n}+1}^{\ast }$ from the baseline distribution $F_{0}$, (\ref%
{baseline}). Equations (\ref{EFgivAlphaBase2})-(\ref{EFgivAlphaBase3}) in
particular show that the conditional posterior predictive distribution
function $\Pr (\mathbf{z}_{n+1}\in B\,|\,\mathbf{Z}_{n},\alpha )$ is a
linear combination of two data sets, namely, the empirical data set $\mathbf{%
Z}_{n}$ with empirical distribution function $\widehat{F}_{n}(B)$, and an
imaginary data set having distribution function $F_{0}$, with sample sizes
(weights) $n$ and $\alpha $ (resp.). We will revisit this point in the next
section.

Finally, the posterior distribution $\Pi (\alpha \,|\,\mathbf{Z}_{n})$ has
p.d.f.: 
\end{subequations}
\begin{equation}
\pi (\alpha \,|\,c_{n})\propto \pi (\alpha )\alpha ^{c_{n}}\Gamma (\alpha
)/\Gamma (\alpha +n),  \label{aPosterior}
\end{equation}%
up to a normalization constant (Nandram \&\ Choi, 2004, p. 828\nocite%
{NandramChoi04}).

\section{Review of OLS properties, and Connections with MDP model}

Now we briefly review of the key properties of the OLS estimator for the
linear model (for more details, see Greene, 2012\nocite{Greene12}). Then we
present connections between the OLS estimator, ridge regression, and
posterior inference with the MDP\ model. This will set up the discussion in
the next section on the MDP-based bootstrap procedure, and the associated
(WLS)\ posterior mean and heteroscedastic-consistent posterior covariance
matrix estimator.

The linear regression equation, defined by $y_{i}=\beta _{1}x_{i1}+\cdots
+\beta _{K}x_{iK}+\varepsilon _{i}=\mathbf{x}_{i}^{\intercal }\boldsymbol{%
\beta }$ for observations indexed by $i=1,\ldots ,n$, has regression errors $%
(\varepsilon _{1},\ldots ,\varepsilon _{i},\ldots ,\varepsilon _{n})$
assuming corresponding variances $\Phi =\mathrm{diag}(\sigma _{1}^{2},\ldots
,\sigma _{n}^{2})$, and assuming exogeneity, i.e., $\mathbb{E}[\varepsilon
_{j}\,|\,\mathbf{x}_{i}]=0,$ for $i,j=1,\ldots ,n$. The OLS estimator of the
coefficients $\boldsymbol{\beta }$ is given by:%
\begin{equation*}
\widehat{\boldsymbol{\beta }}=\widehat{\boldsymbol{\beta }}_{n}=(\mathbf{X}%
^{\intercal }\mathbf{X})^{-1}\mathbf{X}^{\intercal }\mathbf{y}=(\mathbf{X}%
_{c_{n}}^{\ast \intercal }\mathrm{diag}(\mathbf{n}_{c_{n}})\mathbf{X}%
_{c_{n}}^{\ast })^{-1}\mathbf{X}_{c_{n}}^{\ast \intercal }\mathrm{diag}(%
\mathbf{n}_{c_{n}})\mathbf{y}_{c_{n}}^{\ast },
\end{equation*}%
and has sampling covariance matrix:%
\begin{equation}
\mathbb{V}(\widehat{\boldsymbol{\beta }})=(\mathbf{X}^{\intercal }\mathbf{X}%
)^{-1}\mathbf{X}^{\intercal }\Phi \mathbf{X}(\mathbf{X}^{\intercal }\mathbf{X%
})^{-1}.
\end{equation}%
If homoscedasticity holds (i.e., $\sigma ^{2}=\sigma _{1}^{2}=\cdots =\sigma
_{n}^{2}$), then $\Phi =\sigma ^{2}\mathbf{I}_{n}$ and $\mathbb{V}(\widehat{%
\boldsymbol{\beta }})=\sigma ^{2}(\mathbf{X}^{\intercal }\mathbf{X})^{-1}$,
and $\widehat{\sigma }^{2}(\mathbf{X}^{\intercal }\mathbf{X})^{-1}$ provides
a consistent estimator of $\mathbb{V}(\widehat{\boldsymbol{\beta }})$ with $%
\widehat{\sigma }^{2}=(\tfrac{1}{n-K})(\mathbf{y}-\mathbf{X}\widehat{%
\boldsymbol{\beta }})^{\intercal }(\mathbf{y}-\mathbf{X}\widehat{\boldsymbol{%
\beta }})$, but is inconsistent otherwise. The (finite-sample)\
heteroscedastic-consistent (sandwich)\ estimator of $\mathbb{V}(\widehat{%
\boldsymbol{\beta }})$ is given by (White, 1980\nocite{White80b}):%
\begin{equation}
\text{\textrm{HC0}}=(\mathbf{X}^{\intercal }\mathbf{X})^{-1}\mathbf{X}%
^{\intercal }\mathrm{diag}(\widehat{u}_{1}^{2},\ldots ,\widehat{u}_{n}^{2})%
\mathbf{X}(\mathbf{X}^{\intercal }\mathbf{X})^{-1},  \label{HC0}
\end{equation}%
where $\widehat{u}_{i}=\widehat{\varepsilon }_{i}=y_{i}-\mathbf{x}%
_{i}^{\intercal }\widehat{\boldsymbol{\beta }}$ for $i=1,\ldots ,n$, which
reduces to \textrm{HC0 }$=\widehat{\sigma }^{2}(\mathbf{X}^{\intercal }%
\mathbf{X})^{-1}$ under homoscedasticity. This consistency does not rely on
exogeneity. Asymptotically ($n\rightarrow \infty $), $n^{1/2}(\widehat{%
\boldsymbol{\beta }}_{n}-\boldsymbol{\beta })\overset{\mathcal{L}}{%
\rightarrow }\mathrm{N}_{K}(\mathbf{0},n\mathbb{V}(\widehat{\boldsymbol{%
\beta }}))$ in law under mild conditions, and $n$\textrm{HC0} consistently
estimates $n\mathbb{V}(\boldsymbol{\beta })$.

Recall from (\ref{EFgivAlphaBase})-(\ref{EFgivAlphaBase3}) that the
conditional posterior expectation $\mathbb{E}[F(\cdot )\,|\,\mathbf{Z}%
_{n},\alpha ]$ under the MDP model is a linear combination of two
distribution functions, $\widehat{F}_{n}$ and $F_{0}$, corresponding to two
data sets (resp.)\ of total sample size $\alpha +n$. The empirical c.d.f., $%
\widehat{F}_{n}$, which describes the data set $(\mathbf{X}_{n},\mathbf{y}%
_{n})$, has sample mean vector $\widehat{\mathbf{m}}_{\mathbf{z}}=(\widehat{%
\mathbf{m}}_{\mathbf{x}}^{\intercal },\widehat{m}_{Y})^{\intercal }$, and $%
(K+1)\times (K+1)$ covariance matrix $\widehat{\mathbf{V}}_{\mathbf{z}}$,
including the $K\times K$ covariance matrix $\widehat{\mathbf{V}}_{\mathbf{x}%
}$ of $\mathbf{X}$ and the $K\times 1$ vector $\widehat{\mathbf{V}}_{\mathbf{%
x}Y}$ of covariances between the columns of $\mathbf{X}$ and $\mathbf{y}$
(resp.). The MDP\ baseline distribution, $F_{0}$ (in (\ref{baseline})),
which describes the distribution of imaginary data set of $S$ observations,
given by $\overline{\mathbf{Z}}_{S}=(\overline{\mathbf{X}}_{S},\overline{%
\mathbf{y}}_{S})=((\overline{x}_{sk})_{S\times K},(\overline{y}%
_{sk})_{S\times 1})$, has baseline mean $\mathbf{m}_{\mathbf{z}}=(\mathbf{m}%
_{\mathbf{x}}^{\intercal },m_{Y})^{\intercal }$ and $(K+1)\times (K+1)$
covariance matrix $\mathbf{V}_{\mathbf{z}}$, including the $K\times K$
covariance matrix $\mathbf{V}_{\mathbf{x}}$ of $\boldsymbol{X}$ and the $%
K\times 1$ vector $\mathbf{V}_{\mathbf{x}Y}$ of covariances of each column
of $\boldsymbol{X}$ with $Y$ (resp.).

Let $(\mathbb{X},\mathbb{Y})=\left( \QTATOP{\mathbf{X}_{c_{n}}^{\ast }}{%
\overline{\mathbf{X}}_{S}},\QTATOP{\mathbf{y}_{c_{n}}^{\ast }}{\overline{%
\mathbf{y}}_{S}}\right) $. If $F_{0}$ is a general baseline distribution in
the MDP model, perhaps not a ridge baseline (\ref{RidgeBase}), then the OLS
estimator $\widehat{\boldsymbol{\beta }}$ from the data $(\mathbb{X},\mathbb{%
Y})$ satisfies the equalities: 
\begin{subequations}
\label{newOLS}
\begin{eqnarray}
\widehat{\boldsymbol{\beta }} &=&\left[ \left( \QDATOP{\mathbf{X}}{\overline{%
\mathbf{X}}_{S}}\right) ^{\intercal }\mathrm{diag}(\mathbf{1}_{n}^{\intercal
},(\tfrac{\alpha }{S})\mathbf{I}_{S})\left( \QDATOP{\mathbf{X}}{\overline{%
\mathbf{X}}_{S}}\right) \right] ^{-1}\left( \QDATOP{\mathbf{X}}{\overline{%
\mathbf{X}}_{S}}\right) ^{\intercal }\mathrm{diag}(\mathbf{1}_{n}^{\intercal
},(\tfrac{\alpha }{S})\mathbf{I}_{S})\left( \QDATOP{\mathbf{y}}{\overline{%
\mathbf{y}}_{S}}\right)   \label{newOLS0} \\
&=&\left[ \left( \QDATOP{\mathbf{X}_{c_{n}}^{\ast }}{\overline{\mathbf{X}}%
_{S}}\right) ^{\intercal }\mathrm{diag}(\mathbf{n}_{c_{n}}^{\intercal },(%
\tfrac{\alpha }{S})\mathbf{I}_{S})\left( \QDATOP{\mathbf{X}_{c_{n}}^{\ast }}{%
\overline{\mathbf{X}}_{S}}\right) \right] ^{-1}\left( \QDATOP{\mathbf{X}%
_{c_{n}}^{\ast }}{\overline{\mathbf{X}}_{S}}\right) ^{\intercal }\mathrm{diag%
}(\mathbf{n}_{c_{n}}^{\intercal },(\tfrac{\alpha }{S})\mathbf{I}_{S})\left( 
\QDATOP{\mathbf{y}_{c_{n}}^{\ast }}{\overline{\mathbf{y}}_{S}}\right) 
\label{newOLS1} \\
&=&\left( \mathbb{X}^{\intercal }\mathrm{diag}(\mathbf{n}^{\intercal },(%
\tfrac{\alpha }{S})\mathbf{I}_{S})\mathbb{X}\right) ^{-1}\mathbb{X}%
^{\intercal }\mathrm{diag}(\mathbf{n}^{\intercal },(\tfrac{\alpha }{S})%
\mathbf{I}_{S})\mathbb{Y}  \label{newOLS2} \\
&=&\left( \mathbb{X}^{\intercal }[\tfrac{1}{\alpha +n}\mathrm{diag}(\mathbf{n%
}^{\intercal },(\tfrac{\alpha }{S})\mathbf{I}_{S})]\mathbb{X}\right) ^{-1}%
\mathbb{X}^{\intercal }[\tfrac{1}{\alpha +n}\mathrm{diag}(\mathbf{n}%
^{\intercal },(\tfrac{\alpha }{S})\mathbf{I}_{S})]\mathbb{Y}  \label{newOLS3}
\\
&=&\left( n(\widehat{\mathbf{V}}_{\mathbf{x}}+\widehat{\mathbf{m}}_{\mathbf{x%
}}\widehat{\mathbf{m}}_{\mathbf{x}}^{\intercal })+\alpha (\mathbf{V}_{%
\mathbf{x}}+\mathbf{m}_{\mathbf{x}}\mathbf{m}_{\mathbf{x}}^{\intercal
}{})\right) ^{-1}\left( n(\widehat{\mathbf{V}}_{\mathbf{x}Y}+\widehat{m}_{Y}%
\widehat{\mathbf{m}}_{\mathbf{x}})+\alpha (\mathbf{V}_{\mathbf{x}Y}+m_{Y}%
\mathbf{m}_{\mathbf{x}})\right) ,  \label{newOLS4}
\end{eqnarray}%
where in (\ref{newOLS0})-(\ref{newOLS3}), the diagonal elements of $(\alpha
/S)\mathbf{I}_{S}$ sum to $\alpha $, the number of prior imaginary
observations. Also, (\ref{newOLS4}) still yields the OLS estimator $\widehat{%
\boldsymbol{\beta }}$ after replacing $n$ with $\frac{n}{\alpha +n}$, and
replacing $\alpha $ with $\frac{\alpha }{\alpha +n}$. If $\alpha $ is a
positive integer with $S=\alpha $, then $(\alpha /S)\mathbf{I}_{S}=(\alpha
/\alpha )\mathbf{I}_{\alpha }=\mathbf{I}_{\alpha }$, the diagonal elements
of $\mathbf{I}_{\alpha }$ sum to $\alpha $, and the OLS estimator (\ref%
{newOLS}) has the familiar form, given by $\widehat{\boldsymbol{\beta }}=(%
\mathbf{X}_{n+\alpha }^{\intercal }\mathbf{X}_{n+\alpha })^{-1}\mathbf{X}%
_{n+\alpha }^{\intercal }\mathbf{y}_{n+\alpha }$, where $(\mathbf{X}%
_{n+\alpha },\mathbf{y}_{n+\alpha })=\left( \QTATOP{\mathbf{X}_{n}}{%
\overline{\mathbf{X}}_{\alpha }}\QTATOP{\mathbf{y}_{n}}{\overline{\mathbf{y}}%
_{\alpha }}\right) $.

Given any precision parameter $\alpha >0$, where possibly $S\neq \alpha $,
an extension of the fractional imputation procedure (e.g., Kim \&\ Kim, 2012%
\nocite{KimKim12}) can be used to simulate the imaginary data set $(%
\overline{\mathbf{X}},\overline{\mathbf{y}})$. This would involve taking a
large number ($S$) of Monte Carlo sample draws $\mathbf{z}_{c_{n}+1,s}=(%
\mathbf{x}_{c_{n}+1,s},y_{c_{n}+1,s})\sim F_{0}$, for $s=1,\ldots ,S$, to
provide the Monte Carlo estimator $\tfrac{1}{S}\tsum\nolimits_{s=1}^{S}%
\mathbf{z}_{c_{n}+1,s}\mathbf{z}_{c_{n}+1,s}^{\top }=(\widehat{\mathbf{V}_{%
\mathbf{z}}+\mathbf{m}_{\mathbf{z}}\mathbf{m}_{\mathbf{z}}^{\intercal }})$,
including $(\widehat{\mathbf{V}_{\mathbf{x}}+\mathbf{m}_{\mathbf{x}}\mathbf{m%
}_{\mathbf{x}}^{\intercal }{}})$ and $(\widehat{\mathbf{V}_{\mathbf{x}%
Y}+m_{Y}\mathbf{m}_{\mathbf{x}}})$, since p$\lim_{S\rightarrow \infty }%
\tfrac{1}{S}\tsum\nolimits_{s=1}^{S}\mathbf{z}_{c_{n}+1,s}\mathbf{z}%
_{c_{n}+1,s}^{\top }=(\mathbf{V}_{\mathbf{z}}+\mathbf{m}_{\mathbf{z}}\mathbf{%
m}_{\mathbf{z}}^{\intercal })$ by the law of large numbers. Then, the OLS
estimator (\ref{newOLS}) is obtained after plugging in the simulated data $(%
\mathbf{x}_{c_{n}+1,s}^{\intercal })_{s=1}^{S}$ in place of $\overline{%
\mathbf{X}}_{S}$ and plugging in $(y_{c_{n}+1,s})_{s=1}^{S}$ in place of $%
\overline{\mathbf{y}}_{S}$ in (\ref{newOLS1}), using fractional weights $%
(\alpha /S)\mathbf{1}_{S}$; or by plugging in $(\widehat{\mathbf{V}_{\mathbf{%
x}}+\mathbf{m}_{\mathbf{x}}\mathbf{m}_{\mathbf{x}}^{\intercal }{}})$ and $(%
\widehat{\mathbf{V}_{\mathbf{x}Y}+m_{Y}\mathbf{m}_{\mathbf{x}}})$ in place
of $(\mathbf{V}_{\mathbf{x}}+\mathbf{m}_{\mathbf{x}}\mathbf{m}_{\mathbf{x}%
}^{\intercal }{})$ and $(\mathbf{V}_{\mathbf{x}Y}+m_{Y}\mathbf{m}_{\mathbf{x}%
})$ (resp.)\ in (\ref{newOLS4}).

If $F_{0}$ is chosen as the ridge baseline prior (\ref{RidgeBase}), then the
OLS estimator (\ref{newOLS}) is equal to: 
\end{subequations}
\begin{subequations}
\label{ridgeOLS}
\begin{eqnarray}
\widehat{\boldsymbol{\beta }} &=&\left[ \left( \QDATOP{\mathbf{X}%
_{c_{n}}^{\ast }}{\overline{\mathbf{X}}_{K}}\right) ^{\intercal }\mathrm{diag%
}(\mathbf{n}_{c_{n}}^{\intercal },\alpha \mathbf{I}_{K})\left( \QDATOP{%
\mathbf{X}_{c_{n}}^{\ast }}{\overline{\mathbf{X}}_{K}}\right) \right]
^{-1}\left( \QDATOP{\mathbf{X}_{c_{n}}^{\ast }}{\overline{\mathbf{X}}_{K}}%
\right) ^{\intercal }\mathrm{diag}(\mathbf{n}_{c_{n}}^{\intercal },\alpha 
\mathbf{I}_{K})\left( \QDATOP{\mathbf{y}_{c_{n}}^{\ast }}{\overline{\mathbf{y%
}}_{K}}\right)   \label{ridgeOLS1} \\
&=&\left[ \left( \QDATOP{(n_{c}^{1/2}x_{ck}^{\ast })_{c_{n}\times K}}{\alpha
^{1/2}\mathrm{diag}(0,v_{\mathbf{x}2},\ldots ,v_{K})^{1/2}}\right)
^{\intercal }\left( \QDATOP{(n_{c}^{1/2}x_{ck}^{\ast })_{c_{n}\times K}}{%
\alpha ^{1/2}\mathrm{diag}(0,v_{\mathbf{x}2},\ldots ,v_{\mathbf{x}K})^{1/2}}%
\right) \right] ^{-1}\left( \QDATOP{n_{c}^{1/2}\mathbf{y}_{c_{n}}^{\ast }}{%
\mathbf{0}_{K}}\right)   \label{ridgeOLS2} \\
&=&\left[ \left( \QDATOP{\mathbf{X}}{\alpha ^{1/2}\mathrm{diag}(0,v_{\mathbf{%
x}2},\ldots ,v_{K})^{1/2}}\right) ^{\intercal }\left( \QDATOP{\mathbf{X}}{%
\alpha ^{1/2}\mathrm{diag}(0,v_{\mathbf{x}2},\ldots ,v_{K})^{1/2}}\right) %
\right] ^{-1}\left( \QDATOP{\mathbf{y}}{\mathbf{0}_{K}}\right) ,
\label{ridgeOLS3} \\
\overline{\mathbf{X}}_{K} &=&\mathrm{diag}(\mathbf{V}_{\mathbf{x}}+\mathbf{m}%
_{\mathbf{x}}\mathbf{m}_{\mathbf{x}}^{\intercal }{})^{1/2}=\mathrm{diag}%
(0,v_{\mathbf{x}2},\ldots ,v_{\mathbf{x}K})^{1/2}; \\
\overline{\mathbf{y}}_{K} &=&(\mathbf{V}_{\mathbf{x}Y}+m_{Y}\mathbf{m}_{%
\mathbf{x}})^{1/2}=(\mathbf{0}_{K}+\mathbf{0}_{K})^{1/2}=\mathbf{0}_{K},
\end{eqnarray}%
with imaginary data $(\overline{\mathbf{X}}_{K},\overline{\mathbf{y}}_{K})$,
obtained by deterministic single imputation, without any simulation or Monte
Carlo error. Then the OLS estimator (\ref{newOLS}) corresponds to the
generalized ridge regression estimator (Hoerl \&\ Kennard, 1970\nocite%
{HoerlKennard70}) with shrinkage parameters $\alpha \cdot (0,v_{2},\ldots
,v_{K})$. Further, if the unit ridge baseline prior is chosen, with $\mathbf{%
V}_{\mathbf{x}}=(0,\mathbf{1}_{K-1}^{\intercal })$, then $\alpha ^{1/2}%
\overline{\mathbf{X}}_{K}=\alpha ^{1/2}\cdot \{(\mathbf{V}_{\mathbf{x}}+%
\mathbf{m}_{\mathbf{x}}\mathbf{m}_{\mathbf{x}}^{\intercal
}{})\}^{1/2}=\alpha ^{1/2}\mathbf{1}_{K}^{\intercal }$ in terms of (\ref%
{ridgeOLS1})-(\ref{ridgeOLS3}), then the OLS estimator (\ref{newOLS})
coincides with the ridge regression estimator with coefficient shrinkage
parameter $\alpha $ (Hoerl \&\ Kennard, 1970\nocite{HoerlKennard70}), as
Hastie et al. (2009\nocite{HastieTibsFriedman09}, p. 96) observed without
any reference to the DP. Finally, under the ridge baseline prior, we may
just set $(\mathbb{X},\mathbb{Y}):=\left( \QTATOP{\mathbf{X}_{c_{n}}^{\ast }%
}{\overline{\mathbf{X}}_{K}},\QTATOP{\mathbf{y}_{c_{n}}^{\ast }}{\overline{%
\mathbf{y}}_{K}}\right) .$

\section{Bootstrap Approximation to the MDP Posterior Distribution}

Generally speaking, using an MDP model, it is possible to employ a bootstrap
procedure for the inference of a random posterior functional, say $\phi
:=\phi (F)$, having c.d.f. $G(\mathbf{t}\,|\,\mathbf{Z}_{n})=\Pr (\phi
(F)\leq \mathbf{t}\,|\,\mathbf{Z}_{n})$, marginally over the posterior
distribution of $\alpha $. Conditionally on a posterior draw $\alpha \sim
\pi (\alpha \,|\,c_{n})$, this procedure approximates $G(\mathbf{t}\,|\,%
\mathbf{Z}_{n},\alpha )$ by the c.d.f. $G^{\ast }(\mathbf{t}\,|\,\mathbf{Z}%
_{n},\alpha )=\Pr (\phi (F^{\ast })\leq \mathbf{t}\,|\,\mathbf{Z}_{n},\alpha
)$, estimated by $\widehat{G}^{\ast }(\mathbf{t}\,|\,\mathbf{Z}_{n},\alpha )=%
\frac{1}{B}\tsum_{b=1}^{B}\mathbf{1}(\phi (F_{b}^{\ast })\leq \mathbf{t})$
given a large number $B$ of bootstrap samples $\{F_{b}^{\ast }\}_{b=1}^{B}$.
Specifically, each random c.d.f. $F_{b}^{\ast }$ is constructed by: 
\end{subequations}
\begin{equation}
F_{b}^{\ast }(\mathbf{t})\,=\tsum\nolimits_{c=1}^{c_{n}+1}\tfrac{%
n_{cb}^{\ast }}{n+\alpha +1}\mathbf{1}(\mathbf{z}_{c}^{\ast }\leq \mathbf{t}%
)+\tfrac{(\alpha -\mathrm{floor}(\alpha ))}{n+\alpha +1}\mathbf{1}(\mathbf{z}%
_{c_{n}+1+\mathbf{1}(\mathrm{ceil}(\alpha )>\alpha ),b}^{\ast }\leq \mathbf{t%
}),  \label{BootCDF}
\end{equation}%
given $n+\mathrm{ceil}(\alpha )+1$ draws of $\mathbf{z}_{c}^{\ast }$ from
the posterior predictive distribution $\mathbb{E}[F(\cdot )\,|\,\mathbf{Z}%
_{n},\alpha ]=\tsum\nolimits_{c=1}^{c_{n}}\tfrac{n_{c}}{\alpha +n}\delta _{%
\mathbf{z}_{c}^{\ast }}(\cdot )+\tfrac{\alpha }{\alpha +n}F_{0}(\cdot )$
given by (\ref{EFgivAlphaBase3}), and given a draw $\mathbf{n}%
_{c_{n}+1,b}^{\ast }=(n_{cb}^{\ast })_{(c_{n}+1)\times 1}$ from a
multinomial distribution having $n+\mathrm{ceil}(\alpha )+1$ trials and
event probabilities $(\tfrac{n_{1}}{\alpha +n},\ldots ,\tfrac{n_{c}}{\alpha
+n},\ldots ,\tfrac{n_{c_{n}}}{\alpha +n},\tfrac{\alpha }{\alpha +n})$
(Hjort, 1985\nocite{Hjort85}). Here, $\mathrm{floor}(\cdot )$ and $\mathrm{%
ceil}(\cdot )$ refer to the floor and ceiling functions. The last term in (%
\ref{BootCDF}) ensures that the effective number of multinomial trials is $%
n+\alpha +1$, whether or not $\alpha >0$ is a positive integer. If each
bootstrap sample has (effective)\ size $n+\alpha +1$, then the conditional
posterior expectation and variance, $\mathbb{E}[F^{\ast }(B)\,|\,\mathbf{Z}%
_{n},\alpha ]$ and $\mathbb{V}[F^{\ast }(B)\,|\,\mathbf{Z}_{n},\alpha ]$
(for $\forall B\in \mathcal{B}(\mathcal{Z})$) equals to that of (\ref%
{EFgivAlphaBase})-(\ref{EFgivAlphaBase3}) and (\ref{VFgivAlphaBase}) for $F$%
; while $F$ has twice the skewness of $F^{\ast }$ but is small (Hjort, 1985%
\nocite{Hjort85}). Then for well-behaved functionals $\phi :=\phi (F)$, the
posterior distributions of $\phi (F)$ and of $\phi (F^{\ast })\,$ are nearly
equal, i.e., $G(\cdot \,|\,\mathbf{Z}_{n},\alpha )\doteq G^{\ast }(\cdot
\,|\,\mathbf{Z}_{n},\alpha )$, given $\alpha $ (Hjort, 1985\nocite{Hjort85}%
), and $G(\cdot \,|\,\mathbf{Z}_{n})\doteq G^{\ast }(\cdot \,|\,\mathbf{Z}%
_{n})$ marginally over the posterior $\pi (\alpha \,|\,c_{n})$. This
justifies a MDP-based approach to the bootstrap.

Using the MDP-based bootstrap, and extending ideas from Section 3, we
perform inference of the posterior mean and covariance matrix of the random
functional $\phi (F^{\ast }):=\boldsymbol{\beta }(F^{\ast })$, chosen as the
OLS estimator for linear regression. In this case, a bootstrap replication
of the OLS estimator, $\widehat{\boldsymbol{\beta }}(F^{\ast })$, is given
by the following sampling scheme: 
\begin{subequations}
\label{bootScheme1}
\begin{eqnarray}
\widehat{\boldsymbol{\beta }}(F^{\ast }) &=&\widehat{\boldsymbol{\beta }}(%
\mathbf{n}^{\ast })=(\mathbb{X}^{\intercal }\mathbf{\,}\mathrm{diag}(\mathbf{%
n}^{\ast })\mathbb{X}\mathbf{\mathbf{)}}^{-1}\mathbb{X}^{\intercal }\mathbf{%
\,}\mathrm{diag}(\mathbf{n}^{\ast })\mathbb{Y},  \label{WLSboot} \\
\mathbf{n}^{\ast }\, &=&\tfrac{n+\alpha +1}{n+\mathrm{ceil}(\alpha )+1}%
(n_{cb}^{\ast \ast })_{(c_{n}+1)\times 1}  \label{multNorm} \\
(n_{cb}^{\ast \ast })_{(c_{n}+1)\times 1}\,|\,\mathbf{Z}_{n},\alpha &\sim &%
\mathrm{Mu}_{c_{n}+S}(n+\mathrm{ceil}(\alpha )+1;\tfrac{n_{1}}{\alpha +n}%
,\ldots ,\tfrac{n_{c_{n}}}{\alpha +n},\tfrac{\alpha /S}{\alpha +n}\otimes 
\mathbf{1}_{S}^{\intercal }),  \label{Mult nstar} \\
\alpha \,|\,\mathbf{Z}_{n} &\sim &\pi (\alpha \,|\,c_{n}),
\end{eqnarray}%
so that $\widehat{\boldsymbol{\beta }}(F^{\ast })$ is a WLS estimator, with
weights given by the multinomial random variable draw, $\mathbf{n}^{\ast
}=(n_{c}^{\ast })_{c=1}^{c_{n}+S}$, scaled by $\tfrac{n+\alpha +1}{n+\mathrm{%
ceil}(\alpha )+1}$, where $\otimes $ is the Kronecker product operator.
Similarly, Lancaster (2003\nocite{Lancaster03}) showed that a random OLS
functional is also a WLS estimator, in the context of Efron's bootstrap.
Also, the current bootstrap sampling scheme (\ref{bootScheme1}) implicitly
samples from the MDP\ baseline distribution ($F_{0}$) because the bottom $S$
rows of $(\mathbb{X},\mathbb{Y})$ already consist of the imaginary
observations sampled from $F_{0}$ (\ref{baseline}) (see Section 3).

Given $(\mathbf{Z}_{n},\alpha )$, the random variate $\mathbf{n}^{\ast }$ (%
\ref{multNorm}) has mean ($\mathbb{E}$) and covariance matrix ($\mathbb{V}$%
): 
\end{subequations}
\begin{subequations}
\label{Boot E V}
\begin{eqnarray}
\mathbb{E}(\mathbf{n}^{\ast }\,|\,\mathbf{Z}_{n},\alpha ) &=&\overline{%
\mathbf{n}}_{\alpha }^{\ast }=(n+\alpha +1)(\tfrac{n_{1}}{\alpha +n},\ldots ,%
\tfrac{n_{c_{n}}}{\alpha +n},\tfrac{\alpha /S}{\alpha +n}\otimes \mathbf{1}%
_{S}^{\intercal })^{\intercal }{},  \label{En | D, alpha} \\
\mathbb{V}(\mathbf{n}^{\ast }\,|\,\mathbf{Z}_{n},\alpha ) &=&\mathrm{diag}(%
\overline{\mathbf{n}}_{\alpha }^{\ast })-(n+\alpha +1)^{-1}\overline{\mathbf{%
n}}_{\alpha }^{\ast }\overline{\mathbf{n}}_{\alpha }^{\ast \intercal }{}.
\label{Vn | D, alpha}
\end{eqnarray}%
Again, in the case of the ridge baseline prior, we can use $(\mathbb{X},%
\mathbb{Y})=\left( \QTATOP{\mathbf{X}_{c_{n}}^{\ast }}{\mathrm{diag}(0,v_{%
\mathbf{x}2},\ldots ,v_{K})^{1/2}},\QTATOP{\mathbf{y}_{c_{n}}^{\ast }}{%
\mathbf{0}_{K}}\right) $ and use $\tfrac{\alpha }{\alpha +n}\otimes \mathbf{1%
}_{K}^{\intercal }$ in place of $\tfrac{\alpha /S}{\alpha +n}\otimes \mathbf{%
1}_{S}^{\intercal }$, in (\ref{WLSboot}) and (\ref{En | D, alpha}).

Let $A$ be a fine grid of $\alpha $ defined over the support of the prior, $%
\pi (\alpha )$. Then, marginalizing over the posterior $\pi (\alpha
\,|\,c_{n})$ (in (\ref{aPosterior})), and by the total law of probability
for expectations and covariances, the marginal expectation and covariance
matrix can be approximated and rapidly computed by: 
\end{subequations}
\begin{subequations}
\label{PostAvg}
\begin{eqnarray}
\mathbb{E}(\mathbf{n}^{\ast }\,|\,\mathbf{Z}_{n}) &=&\overline{\mathbf{n}}%
^{\ast }\approx \dsum\limits_{\alpha \in A}\mathbb{E}(\mathbf{n}^{\ast }\,|\,%
\mathbf{Z}_{n},\alpha )\pi (\alpha \,|\,c_{n}), \\
\mathbb{V}(\mathbf{n}^{\ast }\,|\,\mathbf{Z}_{n}) &\approx
&\dsum\limits_{\alpha \in A}\mathbb{V}(\mathbf{n}^{\ast }\,|\,\mathbf{Z}%
_{n},\alpha )\pi (\alpha \,|\,c_{n})+\dsum\limits_{\alpha \in A}\mathbb{E}(%
\mathbf{n}^{\ast }\,|\,\mathbf{Z}_{n},\alpha )\{\mathbb{E}(\mathbf{n}^{\ast
}\,|\,\mathbf{Z}_{n},\alpha )\}^{\intercal }\pi (\alpha \,|\,c_{n}) \\
&&-\mathbb{E}(\mathbf{n}^{\ast }\,|\,\mathbf{Z}_{n})\{\mathbb{E}(\mathbf{n}%
^{\ast }\,|\,\mathbf{Z}_{n})\}^{\intercal }.
\end{eqnarray}%
We have found that the quantities above are rather robust to choice of fine
grid $A$. We assume that the values of the grid $A$ are equally-spaced by $%
.005$, with minimum $.005$ and maximum $\xi $.

We now consider a deterministic approach to evaluating the distribution of a
functional (e.g., $\boldsymbol{\beta }^{\ast }(F^{\ast })$) \ of the MDP
posterior, as in previous research on DP functionals (Regazzini, et al. 2002%
\nocite{RegazziniGuglielmiDiNunno02}). Specifically, here we employ the
multivariate delta method to approximate the (MDP\ bootstrap)\ posterior
distribution of $\boldsymbol{\beta }(F^{\ast })$ via a Taylor series
approximation $\boldsymbol{\beta }^{\ast }(\mathbf{n}^{\ast })\approx 
\boldsymbol{\beta }(\mathbf{n}^{\ast })$ of $\boldsymbol{\beta }(\mathbf{n}%
^{\ast })$ around the mean, $\overline{\mathbf{n}}^{\ast }$. With $\tfrac{%
\partial \boldsymbol{\beta }(\mathbf{n})}{\partial \mathbf{n}}$ a $K\times
(c_{n}+S)$ matrix of first derivatives (again, $S=K\ $for the ridge baseline
prior), this Taylor series approximation is given by: 
\end{subequations}
\begin{subequations}
\label{Taylor}
\begin{eqnarray}
\boldsymbol{\beta }^{\ast }(\mathbf{n}^{\ast }) &=&\boldsymbol{\beta }(%
\overline{\mathbf{n}}^{\ast })+\left[ \dfrac{\partial \boldsymbol{\beta }(%
\mathbf{n}^{\ast })}{\partial \mathbf{n}^{\ast }}\right] _{\mathbf{n}=%
\overline{\mathbf{n}}^{\ast }}(\mathbf{n}^{\ast }-\overline{\mathbf{n}}%
^{\ast }) \\
&=&\boldsymbol{\beta }(\overline{\mathbf{n}}^{\ast })+[(\mathbb{Y}-\mathbb{X}%
\boldsymbol{\beta }(\overline{\mathbf{n}}^{\ast }))^{\intercal }\otimes (%
\mathbb{X}^{\intercal }\mathrm{diag}(\overline{\mathbf{n}}^{\ast })\mathbb{X}%
\mathbf{\mathbf{)}}^{-1}\mathbb{X}^{\intercal }]\QDOVERD[ ] {\partial 
\mathrm{vec}\{\mathrm{diag}(\mathbf{n})\}}{\partial (\mathbf{n})^{\intercal }%
}_{\mathbf{n}=\overline{\mathbf{n}}^{\ast }}(\mathbf{n}-\overline{\mathbf{n}}%
^{\ast }) \\
&=&\boldsymbol{\beta }(\overline{\mathbf{n}}^{\ast })+\left[ \mathbf{u}(%
\overline{\mathbf{n}}^{\ast })^{\intercal }\otimes \{\mathbb{X}(\mathbb{X}%
^{\intercal }\mathrm{diag}(\overline{\mathbf{n}}^{\ast })\mathbb{X}%
)^{-1}\}^{\intercal }\right] 
\begin{bmatrix}
\mathbf{e}_{1}\mathbf{e}_{1}^{\intercal } \\ 
\vdots  \\ 
\mathbf{e}_{c_{n}+S}\mathbf{e}_{c_{n}+S}^{\intercal }%
\end{bmatrix}%
(\mathbf{n}-\overline{\mathbf{n}}^{\ast }) \\
&=&\boldsymbol{\beta }(\overline{\mathbf{n}}^{\ast })+\mathbf{R}(\overline{%
\mathbf{n}}^{\ast })^{\intercal }(\mathbf{n}^{\ast }-\overline{\mathbf{n}}%
^{\ast }),
\end{eqnarray}%
which is similar but not identical to Poirier's (2011\nocite{Poirier11}, p.
461) approximation, where: 
\end{subequations}
\begin{subequations}
\begin{eqnarray}
\mathbf{u}(\overline{\mathbf{n}}^{\ast }) &=&\mathbb{Y}-\mathbb{X}%
\boldsymbol{\beta }(\overline{\mathbf{n}}^{\ast });\text{ \ }\mathbf{e}_{c}=(%
\mathbf{1}(c=1),\ldots ,\mathbf{1}(c=c_{n}+S))^{\intercal }; \\
\mathbf{R}(\overline{\mathbf{n}}^{\ast }) &=&(\mathbf{e}_{1}\mathbf{e}%
_{1}^{\intercal },\ldots ,\mathbf{e}_{c_{n}+S}\mathbf{e}_{c_{n}+S}^{%
\intercal })(\mathbf{u}(\overline{\mathbf{n}}^{\ast })\otimes \mathbb{X}(%
\mathbb{X}^{\intercal }\mathrm{diag}(\overline{\mathbf{n}}^{\ast })\mathbb{X}%
)^{-1}) \\
&=&\mathrm{diag}\{\mathbf{u}(\overline{\mathbf{n}}^{\ast })\}\mathbb{X}(%
\mathbb{X}^{\intercal }\mathrm{diag}(\overline{\mathbf{n}}^{\ast })\mathbb{X}%
)^{-1},\text{ is }(c_{n}+S)\times K.
\end{eqnarray}

Then the posterior distribution of $\mathbf{n}^{\ast }$ implies that the
approximation (\ref{Taylor}) has exact posterior mean given by the WLS\
estimator: 
\end{subequations}
\begin{equation}
\mathbb{E}(\boldsymbol{\beta }^{\ast }(\mathbf{n}^{\ast })\,|\,\mathbf{Z}%
_{n})=\boldsymbol{\beta }(\overline{\mathbf{n}}^{\ast })=(\mathbb{X}%
^{\intercal }\mathrm{diag}(\overline{\mathbf{n}}^{\ast })\mathbb{X})^{-1}%
\mathbb{X}^{\intercal }\mathrm{diag}(\overline{\mathbf{n}}^{\ast })\mathbb{Y}%
,  \label{Eb}
\end{equation}%
and exact covariance matrix given by the heteroscedastic-consistent sandwich
estimator for WLS (Greene, 2012, p. 319\nocite{Greene12}): 
\begin{subequations}
\label{Vb}
\begin{eqnarray}
\mathbb{V}(\boldsymbol{\beta }^{\ast }(\mathbf{n}^{\ast })\,|\,\mathbf{Z}%
_{n}) &=&\left[ \tfrac{\partial \boldsymbol{\beta }(\mathbf{n}^{\ast })}{%
\partial \mathbf{n}^{\ast }}\right] _{\mathbf{n}=\overline{\mathbf{n}}^{\ast
}}\mathbb{V}(\mathbf{n}^{\ast }\,|\,\mathbf{Z}_{n})\left[ \tfrac{\partial 
\boldsymbol{\beta }(\mathbf{n}^{\ast })}{\partial \mathbf{n}^{\ast }}\right]
_{\mathbf{n}=\overline{\mathbf{n}}^{\ast }}^{\intercal }=\mathbf{R}(%
\overline{\mathbf{n}}^{\ast })^{\intercal }\mathbb{V}(\mathbf{n}^{\ast }\,|\,%
\mathbf{Z}_{n})\mathbf{R}(\overline{\mathbf{n}}^{\ast }) \\
&=&(\mathbb{X}^{\intercal }\mathrm{diag}(\overline{\mathbf{n}}^{\ast })%
\mathbb{X})^{-1}[\mathbb{X}^{\intercal }\mathrm{diag}(\overline{\mathbf{n}}%
^{\ast }\circ \{\mathbf{u}(\overline{\mathbf{n}}^{\ast })\}^{2})\mathbb{X]}(%
\mathbb{X}^{\intercal }\mathrm{diag}(\overline{\mathbf{n}}^{\ast })\mathbb{X}%
)^{-1},
\end{eqnarray}%
where $\circ $ denotes the Hadamard product operator.\ Then the posterior
variances from this matrix, 
\end{subequations}
\begin{equation*}
\mathrm{diag}\{\mathbb{V}(\boldsymbol{\beta }^{\ast }(\mathbf{n})\,|\,%
\mathbf{Z}_{n})\}=\{\mathbb{V}(\beta _{1}^{\ast }(\mathbf{n}^{\ast })\,|\,%
\mathbf{Z}_{n}),\ldots ,\mathbb{V}(\beta _{K}^{\ast }(\mathbf{n}^{\ast
})\,|\,\mathbf{Z}_{n})\},
\end{equation*}%
provide the asymptotic-consistent 95\%\ posterior credible interval, $\beta
_{k}^{\ast }(\mathbf{n}^{\ast })\pm 1.96\{\mathbb{V}(\beta _{k}^{\ast }(%
\mathbf{n}^{\ast })\,|\,\mathbf{Z}_{n})\}^{1/2}$, respectively for $%
k=1,\ldots ,K$.

For fixed $\alpha $, we can write the expectation (\ref{Eb}) as $\boldsymbol{%
\beta }(\overline{\mathbf{n}}_{\alpha }^{\ast })$, and write the posterior
covariance matrix (\ref{Vb}) as $\mathbb{V}(\boldsymbol{\beta }^{\ast }(%
\mathbf{n}_{\alpha }^{\ast })\,|\,\mathbf{Z}_{n})$.

Suppose that the MDP model (\ref{MDP}) assumes a non-informative DP prior,
defined by the specification $\alpha \rightarrow 0$.\ Also suppose that $%
c_{n}=n$, so that $\overline{\mathbf{n}}=\overline{\mathbf{n}}_{0}=(\mathbf{1%
}_{n}^{\intercal },\mathbf{0}_{S}^{\intercal })^{\intercal }$ with $\mathbf{0%
}_{S}$ a $S\times 1$ vector of zeros ($S=K$ for the ridge baseline prior).
Then the conditional posterior distribution (\ref{DPpostCond1}) is Dirichlet
($\mathrm{Di}$), $\boldsymbol{\theta }_{c_{n}}=(\theta _{1},\ldots ,\theta
_{c_{n}})^{\intercal }\,|\,\mathbf{Z}_{n}\sim \mathrm{Di}_{c_{n}}(\mathbf{n}%
_{c_{n}})$, with support points the $c_{n}$ observed cluster values $\{%
\mathbf{z}_{c}^{\ast }\}_{c=1}^{c_{n}\leq n}$, where $\theta _{c}=\Pr \{%
\mathbf{z}=\mathbf{z}_{c}^{\ast })\}$ for $c=1,\ldots ,c_{n}$, which
coincides with the posterior distribution of sampling probabilities under
the non-informative Bayesian Bootstrap (Rubin, 1981\nocite{Rubin81}). Then
the posterior predictive probability distribution (\ref{EFgivAlphaBase})-(%
\ref{EFgivAlphaBase3}) reduces to $\Pr (\mathbf{z}_{n+1}\in B\,|\,\mathbf{Z}%
_{n})=\widehat{F}_{n}(B)=\tsum\nolimits_{c=1}^{c_{n}}\tfrac{n_{c}}{n}\delta
_{\mathbf{z}_{c}^{\ast }}(B)$, which is the distribution function employed
by the classical (pairs)\ bootstrap (Efron, 1979\nocite{Efron79}); the
posterior mean (\ref{Eb}) is nearly equal to the OLS estimator, with $%
\boldsymbol{\beta }(\overline{\mathbf{n}}_{0})\approx (\mathbf{\mathbf{X}}%
^{\intercal }\mathbf{\mathbf{X}})^{-1}\mathbf{\mathbf{X}}^{\intercal }%
\mathbf{\mathbf{y}}$ and $\overline{\mathbf{n}}_{0}=\lim_{\alpha \rightarrow
0}\overline{\mathbf{n}}_{\alpha }$; and the posterior covariance matrix (\ref%
{Vb}) is nearly equal to the heteroscedasticity consistent (sandwich)\
covariance matrix estimator of White (1980\nocite{White80b}) using total
sample size weight of $(n+1)$, with: 
\begin{equation}
\mathbb{V}(\boldsymbol{\beta }^{\ast }(\mathbf{n}^{\ast })\,|\,\mathbf{Z}%
_{n})=(\mathbf{\mathbf{X}}^{\intercal }\mathbf{\mathbf{X}})^{-1}[\mathbf{%
\mathbf{X}}^{\intercal }\mathrm{diag}(\{\mathbf{u}(\overline{\mathbf{n}}%
_{0}^{\ast })\}^{2})\mathbf{\mathbf{X]}}(\mathbf{\mathbf{X}}^{\intercal }%
\mathbf{\mathbf{X}})^{-1}\approx \text{\textrm{HC0}}.  \label{WhiteCov1}
\end{equation}

\section{Sensitivity and VoE Analysis\ Methods}

We propose and describe sensitivity analysis methods that can be applied in
settings where the assumption of exogeneity may be empirically violated.
Suppose that the following linear regression equation holds true for a given
population: 
\begin{equation}
y_{i}=\beta _{0}+\underline{\mathbf{x}}_{i}^{\intercal }\boldsymbol{\beta }_{%
\underline{\mathbf{x}}}+\beta _{T}t_{i}+\gamma u_{i}+\varepsilon _{i}=%
\mathbb{E}(Y\,|\,\underline{\mathbf{x}}_{i},t_{i},u_{i})+\varepsilon _{i},%
\text{ }i=1,\ldots ,n,  \label{TrueLin}
\end{equation}%
where $\boldsymbol{\beta }=(\beta _{0},\boldsymbol{\beta }_{\underline{%
\mathbf{x}}}^{\intercal },\beta _{T})^{\intercal }$; $\beta _{T}$ is the
true causal effect of a binary (0,1) treatment variable $T$ on $Y$; $\gamma $
is a possibly non-zero effect of $U$ on $Y$; $T$ may be correlated with $U$;
and the $(\underline{\mathbf{x}}_{i},t_{i},u_{i})$ are realizations of the
random variables $\boldsymbol{X}_{i}^{\intercal }=(\underline{\boldsymbol{X}}%
_{i}^{\intercal },T_{i}),$ and $U$ (resp.).

Suppose for this population that the statistician misspecifies (\ref{TrueLin}%
) by the regression equation:%
\begin{equation}
y_{i}=\beta _{0}+\underline{\mathbf{x}}_{i}^{\intercal }\widetilde{%
\boldsymbol{\beta }}_{\underline{\mathbf{x}}}+\widetilde{\beta }%
_{T}t_{i}+\varepsilon _{iU}=\mathbb{E}(Y\,|\,\underline{\mathbf{x}}%
_{i},t_{i})+\varepsilon _{iU},\text{ }i=1,\ldots ,n,  \label{MispecLin}
\end{equation}%
where $U$ is a missing variable, $\varepsilon _{iU}=\gamma u_{i}+\varepsilon
_{i}$ for $i=1,\ldots ,n$. Then $T$ violates the exogeneity assumption
(i.e., is endogenous) if it is correlated with the error term $\varepsilon
_{U}$, making $U$ a source of hidden bias (Rosenbaum, 2002\nocite%
{Rosenbaum02a}); the OLS (WLS)\ estimator of $\widetilde{\beta }_{T}$ is
inconsistent for $\beta _{T}$ (Greene, 2012, p. 259\nocite{Greene12}); and
the coefficients $\widetilde{\boldsymbol{\beta }}=(\widetilde{\beta }_{0},%
\widetilde{\boldsymbol{\beta }}_{\underline{\mathbf{x}}}^{\intercal },%
\widetilde{\beta }_{T})^{\intercal }$ attain the status as pseudo
parameters, having covariance matrix that can still be consistently
estimated by the sandwich estimator.

Assuming no interactions between $(T,U,\underline{\boldsymbol{X}})$, the
relationship between $\beta _{T}$ and $\widetilde{\beta }_{T}$ is given by:%
\begin{equation}
\beta _{T}=\widetilde{\beta }_{T}-\gamma \tint \tint \{u\mathrm{d}%
F_{U}(u\,|\,T=1,\underline{\mathbf{x}})-u\mathrm{d}F_{U}(u\,|\,T=0,%
\underline{\mathbf{x}})\}\mathrm{d}F_{\underline{\boldsymbol{X}}}(\underline{%
\mathbf{x}}),  \label{betaEqgeneral}
\end{equation}%
where $\mu _{1}$ and $\mu _{0}$ is the mean under distribution (c.d.f.)\ $%
F_{U}(u\,|\,T=1,\underline{\mathbf{x}})$ and $F_{U}(u\,|\,T=0,\underline{%
\mathbf{x}})$ (resp.); and further, if $\mathbb{E}(U\,|\,\underline{\mathbf{x%
}},t)=\mu _{t,\underline{\mathbf{x}}}=\mu _{t}+q(\underline{\mathbf{x}})$
for some function $q$, then $\beta _{T}=\widetilde{\beta }_{T}-\gamma (\mu
_{1}-\mu _{0})$ (VanderWeele \& Arah, 2011\nocite{VanderweeleArah11},
Appendix). Also, the missing variable, $U$, may be assumed to have a
binomial distribution $F_{U}(u\,|\,t,\underline{\mathbf{x}})$ with success
probability $\mu _{t,\underline{\mathbf{x}}}$, with no loss of generality
(Wang \&\ Kreiger, 2006\nocite{WangKrieger06}).

Along these lines, a new Vibration of Effects (VoE) analysis method,
described here, can also be employed for sensitivity analysis. Specifically,
this method provides a way to evaluate how the heteroscedatic-consistent
effect size of the treatment variable, given by $\widetilde{ES}_{T\alpha }=%
\widetilde{\beta }_{T}^{\ast }/\mathbb{V}_{T\alpha }^{1/2}=\widetilde{\beta }%
_{T}^{\ast }(\overline{\mathbf{n}}_{\alpha }^{\ast })/\{\mathbb{V}(\beta
_{T}^{\ast }(\mathbf{n}_{\alpha }^{\ast })\,|\,\mathbf{Z}_{n},\alpha
)\}^{1/2}$, varies as a function of the other covariates that are included
in the regression model, and $\alpha $. To explain this method, assume for
the MDP\ model the ridge baseline prior (\ref{RidgeBase}) with prior
covariances $\mathbb{V}=\mathrm{diag}(0,\mathbf{1}_{K-1}^{\intercal },0)$,
with little loss of generality. Then, given $\alpha $, and after rescaling
each of the columns of $\left[ \left( \QTATOP{\mathbf{X}_{0}}{\sqrt{\alpha 
\mathbf{I}_{K}}}\right) ,\left( \QTATOP{\mathbf{y}}{\mathbf{0}_{K}}\right) %
\right] $ to have mean zero and variance 1, yielding $(\widetilde{\mathbb{X}}%
_{\alpha },\widetilde{\mathbb{Y}})$, where $\mathbf{X}_{0}$ is $\mathbf{X}$
after removing the first column of 1s, the LARS algorithm is run on $(%
\widetilde{\mathbb{X}}_{\alpha },\widetilde{\mathbb{Y}})$ in order to obtain
a sequence of estimates $\widehat{\boldsymbol{\beta }}_{0,\mathrm{lar}%
}^{(\alpha )},\widehat{\boldsymbol{\beta }}_{1,\mathrm{lar}}^{(\alpha )},...,%
\widehat{\boldsymbol{\beta }}_{k,\mathrm{lar}}^{(\alpha )},...,\widehat{%
\boldsymbol{\beta }}_{K-1,\mathrm{lar}}^{(\alpha )}$, where for $k=0,\ldots
,K-2$, $\widehat{\boldsymbol{\beta }}_{k,\mathrm{lar}}^{(\alpha )}$ is the
LARS\ estimate (\ref{newOLS}) of the coefficients that contains the best $k$
out of the total $K-1$ covariates in the regression equation, given $\alpha $%
. Then for each subset $\mathcal{S}_{Tl}^{(\alpha )}$ (for $l=1,2,\ldots $)
of the $K$ covariate subsets that includes the treatment variable $T$, and
now using $(\mathbb{X},\mathbb{Y})_{\alpha }=\left[ \left( \QTATOP{\mathbf{X}%
_{c_{n}}^{\ast }}{\mathbf{0}_{K},\sqrt{\alpha \mathbf{I}_{K}}}\right)
,\left( \QTATOP{\mathbf{y}_{c_{n}}^{\ast }}{\mathbf{0}_{K}}\right) \right] $%
, we compute the WLS estimate $\boldsymbol{\beta }(\overline{\mathbf{n}}%
_{\alpha }^{\ast })$ and the heteroscedastic consistent covariance matrix $%
\mathbb{V}(\boldsymbol{\beta }^{\ast }(\mathbf{n}^{\ast })\,|\,\mathbf{Z}%
_{n},\alpha )$, using (\ref{Eb}) and (\ref{Vb}) (rep.), in order to obtain
the effect size estimate $\widetilde{ES}_{T\alpha }(\mathcal{S}%
_{Tl}^{(\alpha )})$, for $l=1,2,\ldots $. This procedure involving LARS and
subsequent $\widetilde{ES}_{T\alpha }(\mathcal{S}_{Tl}^{(\alpha )})$
estimation, for each covariate subset $\mathcal{S}_{Tl}^{(\alpha )}$, is run
for each value of $\alpha $ over a fine grid $A$ of values in the support of
the prior $\pi (\alpha ).$ The entire procedure yields a large collection of
effect size $\widetilde{ES}_{T\alpha }(\mathcal{S}_{Tl}^{(\alpha )})$
statistics over all relevant covariate subsets $\mathcal{S}_{Tl}^{(\alpha )}$
given $\alpha $, over the grid $A$ of $\alpha $ values, to provide a VoE
analysis of the heteroscedastic-consistent effect size, $\widetilde{ES}_{T}$%
. These effect sizes $\widetilde{ES}_{T\alpha }(\mathcal{S}_{Tl}^{(\alpha )})
$ can be associated with values of the Generalized Information\ Criterion, $%
GIC_{2}(\alpha ,p)=\frac{1}{(n+\alpha )}\{||\mathbb{Y}-\mathbb{X}\boldsymbol{%
\beta }(\overline{\mathbf{n}}_{\alpha }^{\ast })||^{2}+2p\}$ (Fan \&\ Tang,
2013\nocite{FanTang13}), indicating the quality of the predictive fit of the
regression that includes $p$ covariates and penalty $2p$. Effect sizes
associated with smaller values of $GIC_{2}$ may then receive higher priority
for statistical inference.

Moreover, using (\ref{betaEqgeneral}), and a binary missing confounding
variable $U$, we may additionally compute and observe the effect size
estimator $ES_{T\alpha }(\gamma ,\boldsymbol{\lambda })=\beta _{T}/\mathbb{V}%
_{T\alpha }^{1/2}$, over independent standard normal random samples of $%
(\gamma ,\boldsymbol{\lambda })$, where $F_{U}(u\,|\,T=1,\underline{\mathbf{x%
}}\mathbf{;}\boldsymbol{\lambda })$ and $F_{U}(u\,|\,T=0,\underline{\mathbf{x%
}}\mathbf{;}\boldsymbol{\lambda })$ are specified by a binary logistic
regression of $U$ on $(T,\underline{\boldsymbol{X}})$ with coefficients $%
\boldsymbol{\lambda }$. (If the observations of $(T,\underline{\boldsymbol{X}%
}^{\intercal })$ were zero-mean centered before WLS estimation, then these
two binary regressions would be performed conditionally on the maximum and
minimum values of the zero-centered $T$, resp.). Section 7 illustrates this
entire VoE method through the analysis of two real data sets.

\section{Simulation Study}

A simulation study was performed to compare the coverage rates of the 95\%\
heteroscedastic-consistent posterior intervals of the coefficient of a
covariate, obtained from three models (resp.). They include the MDP model
specified under a uniform $\mathrm{un}(\alpha \,|\,0,3)$ prior, the MDP
model under the $\xi $-truncated Cauchy-type prior for $\alpha $, with $\xi
=3$ (Section 2); and the linear model estimated under OLS\ using White's
original sandwich covariance estimator (\textrm{HC0})\ (\ref{HC0}). Also,
for each MDP\ model, we assumed the unit ridge baseline prior. Then, $\alpha 
$ is the coefficient shrinkage penalty parameter of ordinary ridge
regression (Section 3), and the standard \textrm{HC0} model assumes $\alpha
=0$ (Section 4). Before fitting each model to each simulated data set, the
covariate data were zero-mean centered.

\begin{table}[H] \centering%
\begin{tabular}{|ll|ccc|ccc|ccc|ccc|}
\hline
&  & \multicolumn{12}{c|}{Heteroscedasticity Level} \\ \cline{3-14}
&  & \multicolumn{3}{|c|}{$a_{h}=0$ ($0$)} & \multicolumn{3}{|c|}{$a_{h}=1$ (%
$.05$)} & \multicolumn{3}{|c|}{$a_{h}=2$ ($.1$)} & \multicolumn{3}{|c|}{$%
a_{h}=2.5$ ($.15$)} \\ \cline{3-14}
X dist. & \multicolumn{1}{c|}{$n$} & c & u & h & c & u & h & c & u & h & c & 
u & h \\ \hline
U(0,1) & \multicolumn{1}{c|}{10} & $.59$ & $.58$ & $.83$ & $.72$ & $.71$ & $%
.82$ & $.87$ & $.87$ & $.79$ & $.93$ & $.92$ & $.78$ \\ 
Ex(1) & \multicolumn{1}{c|}{10} & $.78$ & $.78$ & $.79$ & $.74$ & $.74$ & $%
.73$ & $.70$ & $.69$ & $.68$ & $.66$ & $.66$ & $.64$ \\ 
N(0,25) & \multicolumn{1}{c|}{10} & $.81$ & $.81$ & $.82$ & $.65$ & $.65$ & $%
.67$ & $.64$ & $.64$ & $.67$ & $.66$ & $.66$ & $.70$ \\ 
AR(1) & \multicolumn{1}{c|}{10} & $.82$ & $.82$ & $.81$ & $.81$ & $.80$ & $%
.80$ & $.78$ & $.79$ & $.78$ & $.77$ & $.77$ & $.77$ \\ \hline
U(0,1) & \multicolumn{1}{c|}{20} & $.74$ & $.74$ & $.90$ & $.84$ & $.84$ & $%
.89$ & $.92$ & $.92$ & $.87$ & $.95$ & $.95$ & $.86$ \\ 
Ex(1) & \multicolumn{1}{c|}{20} & $.86$ & $.86$ & $.86$ & $.81$ & $.81$ & $%
.80$ & $.75$ & $.75$ & $.75$ & $.71$ & $.71$ & $.70$ \\ 
N(0,25) & \multicolumn{1}{c|}{20} & $.88$ & $.88$ & $.88$ & $.84$ & $.84$ & $%
.85$ & $.88$ & $.88$ & $.89$ & $.90$ & $.90$ & $.91$ \\ 
AR(1) & \multicolumn{1}{c|}{20} & $.88$ & $.88$ & $.89$ & $.87$ & $.87$ & $%
.87$ & $.86$ & $.86$ & $.86$ & $.85$ & $.85$ & $.86$ \\ \hline
U(0,1) & \multicolumn{1}{c|}{50} & $.88$ & $.88$ & $.93$ & $.92$ & $.92$ & $%
.93$ & $.94$ & $.94$ & $.92$ & $.94$ & $.94$ & $.92$ \\ 
Ex(1) & \multicolumn{1}{c|}{50} & $.90$ & $.90$ & $.91$ & $.86$ & $.86$ & $%
.86$ & $.82$ & $.82$ & $.82$ & $.82$ & $.82$ & $.82$ \\ 
N(0,25) & \multicolumn{1}{c|}{50} & $.92$ & $.92$ & $.92$ & $.93$ & $.93$ & $%
.93$ & $.96$ & $.96$ & $.97$ & $.98$ & $.98$ & $.98$ \\ 
AR(1) & \multicolumn{1}{c|}{50} & $.93$ & $.93$ & $.93$ & $.92$ & $.92$ & $%
.92$ & $.91$ & $.91$ & $.92$ & $.91$ & $.91$ & $.92$ \\ 
\cline{1-4}\cline{1-9}\cline{7-14}
U(0,1) & \multicolumn{1}{c|}{100} & $.92$ & $.92$ & $.94$ & $.94$ & $.94$ & $%
.94$ & $.94$ & $.94$ & $.94$ & $.94$ & $.94$ & $.94$ \\ 
Ex(1) & \multicolumn{1}{c|}{100} & $.92$ & $.92$ & $.92$ & $.89$ & $.89$ & $%
.90$ & $.89$ & $.89$ & $.89$ & $.90$ & $.90$ & $.90$ \\ 
N(0,25) & \multicolumn{1}{c|}{100} & $.94$ & $.94$ & $.94$ & $.96$ & $.96$ & 
$.96$ & $.98$ & $.98$ & $.98$ & $.99$ & $.99$ & $.99$ \\ 
AR(1) & \multicolumn{1}{c|}{100} & $.94$ & $.94$ & $.94$ & $.93$ & $.93$ & $%
.94$ & $.93$ & $.93$ & $.93$ & $.93$ & $.93$ & $.93$ \\ \hline
U(0,1) & \multicolumn{1}{c|}{500} & $.94$ & $.94$ & $.95$ & $.95$ & $.95$ & $%
.95$ & $.95$ & $.95$ & $.95$ & $.94$ & $.94$ & $.95$ \\ 
Ex(1) & \multicolumn{1}{c|}{500} & $.95$ & $.95$ & $.95$ & $.93$ & $.93$ & $%
.93$ & $.95$ & $.95$ & $.95$ & $.96$ & $.96$ & $.96$ \\ 
N(0,25) & \multicolumn{1}{c|}{500} & $.95$ & $.95$ & $.95$ & $.98$ & $.98$ & 
$.98$ & $1.0$ & $1.0$ & $1.0$ & $1.0$ & $1.0$ & $1.0$ \\ 
AR(1) & \multicolumn{1}{c|}{500} & $.95$ & $.95$ & $.95$ & $.95$ & $.95$ & $%
.95$ & $.95$ & $.95$ & $.95$ & $.95$ & $.95$ & $.95$ \\ \hline
\end{tabular}%
\caption{For the 80 simulation conditions, coverage rates of 95 percent posterior (confidence) interval for: c = MDP with Cauchy type prior; u = MDP with uniform prior; and h = HC0. Heteroscedasticity levels 1, 2, 3, and 4 refer to a = 0, 1, 2, 2.5 for the U(0,1) covariate distribution (resp.), and refer to a = 0, .05,  .1, .15 otherwise (resp.).}%
\label{Table1}%
\end{table}%

The simulation study employed a $4\times 4\times 5$ design that reflects a
wide range of conditions that has been considered in past related research.
Each of the 80 total cells of the design used 10K simulated data sets, for a
total of 800K. Each data set was simulated by taking $n$ samples from the
normal linear model, $y_{i}\,|\,x_{i},\boldsymbol{\beta },\sigma
_{i}^{2}\sim \mathrm{N}(\beta _{0}+\beta _{1}x_{i},\sigma _{i}^{2})$, with $%
\beta _{0}=\beta _{1}=1$ and $\sigma _{i}^{2}=\exp
(a_{h}x_{i}+a_{h}x_{i}^{2})$, for $i=1,\ldots ,n$ (as in Cribari-Neto et
al., 2000\nocite{CribariNetoEtAl00}), according to one of four covariate $%
x_{i}$ sampling distributions$_{i}$; one of four levels $a_{h}$ of
heteroscedasticity; and one of five sample sizes, $n=10,20,50,100,$ and $500$%
. The four covariate distributions are given by the standard uniform
distribution $\mathrm{U}(0,1)$ (Cribari-Neto et al., 2000\nocite%
{CribariNetoEtAl00}), the normal $\mathrm{N}(0,25)$ distribution (Cameron
\&\ Trivedi, 2005, p. 84\nocite{CameronTrivedi05}), the exponential $\mathrm{%
Ex}(1)$ distribution (Hoff \&\ Wakefield, 2013\nocite{HoffWakefield13}), and
the order-1 auto-regressive AR(1)\ model with Student errors (Hansen, 2007%
\nocite{Hansen07}), i.e., $x_{i}=1+.5x_{i-1}+\varepsilon _{i}$, $\varepsilon
_{i}\sim $ \textrm{St}$(0,1,n-1)$, for $i=1,\ldots ,n$. The four
heteroscedasticity levels are given by $a_{h}=0,1,2,2.5$ for the $\mathrm{U}%
(0,1)$ covariate distribution, and otherwise given by $a_{h}=0,.05,.1,.15$
(Cribari-Neto et al., 2000\nocite{CribariNetoEtAl00}), where in each case $%
a_{h}=0$ refers to a condition of homoscedasticity.

Table 1 presents the coverage rates of the 95\%\ heteroscedastic-consistent
posterior intervals for the true data-generating slope coefficient ($\beta
_{1}$), for each of the 80 cells and the three models. Each rate shown is
the proportion of times a model's estimated 95\%\ interval contained the
true data-generating slope value ($\beta _{1}=1$) over the 10K simulated
data sets. Table 2 summarizes the coverage rates of Table 1 by averages and
standard deviations, stratified by covariate distribution,
heteroscedasticity level, and sample size condition. Both tables show that
the coverage rates are generally similar across the three models, especially
for the larger sample size conditions, where the coverage rates for all
three models approach .95. As Table 2 shows for the uniform $\mathrm{U}(0,1)$
covariate distribution, HC0 tended to be closer to .95 for the lower two
heteroscedasticity levels, whereas the converse was true for the higher two
heteroscedasticity levels. The same was true for the $n=10$ and $n=20$
sample size conditions.

However, recall that the simulation study focused on the generation of
positive-definite covariate design matrices ($\mathbf{X}$). A design matrix
that has multicollinearity or is singular can yield an infinite value of 
\textrm{HC0}, whereas for a MDP model with ridge baseline prior will still
yield a defined posterior covariance matrix. This is known to be an
advantage of ridge regression over OLS regression.

\noindent 
\begin{table}[tbp] \centering%
\begin{tabular}{|c|ccc|ccc|ccc|ccc|}
\hline
& \multicolumn{12}{c|}{Heteroscedasticity Level} \\ \cline{2-13}
& \multicolumn{3}{|c|}{$a_{h}=0$ ($0$)} & \multicolumn{3}{|c|}{$a_{h}=1$ ($%
.05$)} & \multicolumn{3}{|c|}{$a_{h}=2$ ($.1$)} & \multicolumn{3}{|c|}{$%
a_{h}=2.5$ ($.15$)} \\ \cline{2-13}
& c & u & h & c & u & h & c & u & h & c & u & h \\ \hline
\multicolumn{1}{|r|}{U(0,1)} & $.81$ & $.81$ & $.91$ & $.87$ & $.87$ & $.91$
& $.92$ & $.92$ & $.89$ & $.94$ & $.94$ & $.89$ \\ 
\multicolumn{1}{|r|}{} & $(.13)$ & $(.14)$ & $(.04)$ & $(.09)$ & $(.09)$ & $%
(.05)$ & $(.03)$ & $(.03)$ & $(.06)$ & $(.01)$ & $(.01)$ & $(.06)$ \\ 
\multicolumn{1}{|r|}{Ex(1)} & $.88$ & $.88$ & $.89$ & $.85$ & $.85$ & $.84$
& $.82$ & $.82$ & $.82$ & $.81$ & $.81$ & $.80$ \\ 
\multicolumn{1}{|r|}{} & $(.06)$ & $(.06)$ & $(.06)$ & $(.07)$ & $(.07)$ & $%
(.07)$ & $(.09)$ & $(.09)$ & $(.09)$ & $(.11)$ & $(.11)$ & $(.12)$ \\ 
\multicolumn{1}{|r|}{N(0,25)} & $.90$ & $.90$ & $.90$ & $.87$ & $.87$ & $.88$
& $.89$ & $.89$ & $.90$ & $.91$ & $.91$ & $.92$ \\ 
\multicolumn{1}{|r|}{} & $(.05)$ & $(.05)$ & $(.05)$ & $(.12)$ & $(.12)$ & $%
(.11)$ & $(.13)$ & $(.13)$ & $(.12)$ & $(.13)$ & $(.13)$ & $(.11)$ \\ 
\multicolumn{1}{|r|}{AR(1)} & $.90$ & $.90$ & $.90$ & $.90$ & $.90$ & $.90$
& $.89$ & $.89$ & $.89$ & $.88$ & $.88$ & $.88$ \\ 
\multicolumn{1}{|r|}{} & $(.05)$ & $(.05)$ & $(.05)$ & $(.05)$ & $(.05)$ & $%
(.05)$ & $(.06)$ & $(.06)$ & $(.06)$ & $(.07)$ & $(.07)$ & $(.07)$ \\ \hline
\multicolumn{1}{|r|}{$n=$ 10} & $.75$ & $.75$ & $.81$ & $.73$ & $.72$ & $.76$
& $.75$ & $.75$ & $.73$ & $.75$ & $.75$ & $.72$ \\ 
\multicolumn{1}{|r|}{} & $(.09)$ & $(.10)$ & $(.01)$ & $(.06)$ & $(.06)$ & $%
(.06)$ & $(.09)$ & $(.09)$ & $(.06)$ & $(.11)$ & $(.11)$ & $(.06)$ \\ 
\multicolumn{1}{|r|}{$n=$ 20} & $.84$ & $.84$ & $.88$ & $.84$ & $.84$ & $.85$
& $.85$ & $.85$ & $.84$ & $.85$ & $.85$ & $.83$ \\ 
\multicolumn{1}{|r|}{} & $(.06)$ & $(.06)$ & $(.01)$ & $(.02)$ & $(.02)$ & $%
(.03)$ & $(.06)$ & $(.06)$ & $(.06)$ & $(.09)$ & $(.09)$ & $(.08)$ \\ 
\multicolumn{1}{|r|}{$n=$ 50} & $.91$ & $.91$ & $.92$ & $.91$ & $.91$ & $.91$
& $.91$ & $.91$ & $.91$ & $.91$ & $.91$ & $.91$ \\ 
\multicolumn{1}{|r|}{} & $(.02)$ & $(.02)$ & $(.01)$ & $(.03)$ & $(.03)$ & $%
(.03)$ & $(.05)$ & $(.05)$ & $(.05)$ & $(.06)$ & $(.06)$ & $(.06)$ \\ 
\multicolumn{1}{|r|}{$n=$ 100} & $.93$ & $.93$ & $.94$ & $.93$ & $.93$ & $%
.93 $ & $.94$ & $.94$ & $.94$ & $.94$ & $.94$ & $.94$ \\ 
\multicolumn{1}{|r|}{} & $(.01)$ & $(.01)$ & $(.01)$ & $(.02)$ & $(.02)$ & $%
(.02)$ & $(.03)$ & $(.03)$ & $(.03)$ & $(.03)$ & $(.03)$ & $(.03)$ \\ 
\multicolumn{1}{|r|}{$n=$ 500} & $.95$ & $.95$ & $.95$ & $.95$ & $.95$ & $%
.95 $ & $.96$ & $.96$ & $.96$ & $.96$ & $.96$ & $.96$ \\ 
\multicolumn{1}{|l|}{} & $(.00)$ & $(.00)$ & $(.00)$ & $(.02)$ & $(.02)$ & $%
(.02)$ & $(.02)$ & $(.02)$ & $(.02)$ & $(.02)$ & $(.02)$ & $(.02)$ \\ \hline
Total & $.87$ & $.87$ & $.90$ & $.87$ & $.87$ & $.88$ & $.88$ & $.88$ & $.88$
& $.88$ & $.88$ & $.87$ \\ 
& $(.09)$ & $(.09)$ & $(.05)$ & $(.09)$ & $(.09)$ & $(.08)$ & $(.10)$ & $%
(.10)$ & $(.09)$ & $(.10)$ & $(.10)$ & $(.10)$ \\ \hline
\end{tabular}%
\caption{Means (standard deviations) of the simulation results of Table 1.}%
\label{Table2}%
\end{table}%

\section{Real Data Applications}

We now illustrate the application of the MDP model on two real data sets,
assuming unit ridge baseline prior, and a uniform $\mathrm{un}(\alpha
\,|\,0,3)$ prior for $\alpha $.

\subsection{LMT\ Data}

Here we analyze real data set of observations from $n=347$ undergraduate
teacher education students ($89.9\%$ female) who each attended one of four
Chicago universities between the Fall 2007 semester and Fall 2013 spring
semesters, inclusive, excluding summers. The primary aim of the analysis is
to infer the effect of the new teacher education curriculum (versus old
curriculum)\ on a dependent variable ($Y$) of math teaching ability. Here,
ability is defined as the number-correct score obtained on a 25-item test of
Learning Math for Teaching (LMT, 2012\nocite{LMT12}), after completing a
course on algebra teaching. Three covariates were considered, namely, Year
and Year$^{2}$, and CTPP $=\mathbf{1}(\mathrm{Year}\geq 2010.9)$, an
indicator of the administration of the new (versus old)\ teaching
curriculum. All covariates were rescaled to have mean zero and variance 1
before data analysis.

\noindent 
\begin{table}[H] \centering%
\begin{tabular}{ccccc|cc}
\cline{2-7}
& $\boldsymbol{\beta }(\overline{\mathbf{n}}^{\ast })$ & pSD & ES & 95\%PI $%
\boldsymbol{\beta }(\overline{\mathbf{n}}^{\ast })$ & OLS$\ \widehat{%
\boldsymbol{\beta }}$ & SE \\ \hline
\multicolumn{1}{r}{Intercept} & $12.90$ & $.18$ & $72.76$ & $(12.55,13.24)$
& $12.90$ & $.18$ \\ 
\multicolumn{1}{r}{Year} & $-.69$ & $.57$ & $-1.21$ & $(-1.81,.43)$ & $%
-478.17$ & $426.54$ \\ 
\multicolumn{1}{r}{Year$^{2}$} & $-.65$ & $.57$ & $-1.14$ & $(-1.78,.47)$ & $%
476.75$ & $426.54$ \\ 
\multicolumn{1}{r}{CTPP} & $.60$ & $.28$ & $2.\allowbreak 10$ & $(.04,1.15)$
& $.67$ & $.28$ \\ \hline
\end{tabular}%
\caption{The slope coefficient estimates for the real data set, including heteroscedastic-consistent 
posterior standard deviation (pSD) and robust 95 percent posterior intervals (PI). Also included are the  OLS estimates of the coefficients and their respective robust standard errors (SE).}%
\label{Table3}%
\end{table}%

\begin{center}
------------------

Figure 1

------------------
\end{center}

Table 3 presents the results of the data analysis, in terms of the MDP-based
WLS\ estimates ($\boldsymbol{\beta }(\overline{\mathbf{n}}^{\ast })$) and
their respective heteroscedastic-consistent (robust)\ 95\% posterior
credible intervals. The CTPP\ causal effect was significant, as this
covariate's 95\%\ heteroscedastic-consistent posterior interval $(.04,1.15)$
excludes zero. This table also presents the OLS\ estimates ($\widehat{%
\boldsymbol{\beta }}$) and their respective robust standard errors based on
the ordinary sandwich estimator, and show that the OLS estimate of the Year
slope coefficient and its standard error are large (in absolute value) due
to the multicollinearity of the Year and Year$^{2}$ covariate observations.
This is not true for any of the WLS\ estimates and corresponding posterior
standard deviations (pSD). As mentioned, multicollinearity can explode the
variance of OLS estimates. In contrast, in ridge regression, provided by the
MDP ridge baseline prior, helps control the size of the WLS and variance
estimates of the coefficients by shrinking coefficient estimates towards
zero.

Figure 1 presents the results of the VoE analysis, relating the CTPP effect
size, $GIC_{2}$, $\alpha $, and subsets of the covariates (Year,Year$^{2}$%
,CTPP) chosen by the LARS algorithm, only among the subsets that included
CTPP. These results are based on a total of 605 regressions (CTPP\ effect
sizes). Over these conditions, the CTPP effect is rather stable. The figure
also presents a sensitivity analysis of a hypothetical missing variable $U$,
over 50 independent standard normal random samples of $(\gamma ,\boldsymbol{%
\lambda })$, and shows some instability of the CTPP\ effect size with
respect to this variable.

\subsection{PIRLS\ Data}

A data set was obtained from the 2006 Progress in International Reading
Literacy Study (PIRLS), on 565 low-income students from 21 U.S. elementary
schools. For data analysis, the dependent variable is student literacy score
(zREAD), along with 8 covariates:\ student male status (1 if MALE, or 0),
AGE, class size (SIZE), class percent of English language learners (ELL);
teacher years of experience (TEXP4) and education level (EDLEVEL = 5 if
bachelor's; EDLEVEL = 6 if at least master's degree); school enrollment
(ENROL) and safety rating (SAFE = 1 is high; SAFE = 3 is low). Each variable
in the data set was rescaled to z-scores having mean 0 and variance 1.

\begin{table}[H] \centering%
\begin{tabular}{ccccc|cc}
\cline{2-7}
& $\boldsymbol{\beta }(\overline{\mathbf{n}}^{\ast })$ & pSD & ES & 95\%PI $%
\boldsymbol{\beta }(\overline{\mathbf{n}}^{\ast })$ & OLS$\ \widehat{%
\boldsymbol{\beta }}$ & SE \\ \hline
\multicolumn{1}{r}{Intercept} & $-.48$ & $.04$ & $-12.67$ & $(-.56,-.41)$ & $%
-.48$ & $.04$ \\ 
\multicolumn{1}{r}{MALE} & $-.06$ & $.04$ & $-1.\allowbreak 44$ & $(-.13,.02)
$ & $-.06$ & $.04$ \\ 
\multicolumn{1}{r}{AGE} & $-.20$ & $.04$ & $-5.22$ & $(-.27,-.12)$ & $-.20$
& $.04$ \\ 
\multicolumn{1}{r}{SIZE} & $-.07$ & $.05$ & $-1.\allowbreak 48$ & $(-.16,.02)
$ & $-.07$ & $.05$ \\ 
\multicolumn{1}{r}{ELL} & $-.13$ & $.05$ & $-2.\allowbreak 81$ & $(-.22,-.04)
$ & $-.13$ & $.05$ \\ 
\multicolumn{1}{r}{TEXP4} & $.14$ & $.04$ & $3.\allowbreak 23$ & $(.05,.22)$
& $.14$ & $.04$ \\ 
EDLEVEL & $.03$ & $.04$ & $.77$ & $(-.05,.11)$ & $.03$ & $.04$ \\ 
\multicolumn{1}{r}{ENROL} & $.30$ & $.04$ & $7.\allowbreak 60$ & $(.22,.37)$
& $.30$ & $.04$ \\ 
\multicolumn{1}{r}{SAFE} & $-.16$ & $.04$ & $-3.71$ & $(-.24,-.07)$ & $-.16$
& $.04$ \\ \hline
\end{tabular}%
\caption{The slope coefficient estimates for the real data set, including heteroscedastic-consistent 
posterior standard deviation (pSD) and robust 95 percent posterior intervals (PI). Also included are the  OLS estimates of the coefficients and their respective robust standard errors (SE).}%
\label{Table4}%
\end{table}%

Table 4 presents the results of the data analysis, including the MDP-based
WLS\ estimates ($\boldsymbol{\beta }(\overline{\mathbf{n}}^{\ast })$), their
respective heteroscedastic-consistent (robust)\ 95\% posterior credible
intervals. According to the MDP\ model, teacher's years of experience
(TEXP4) was found to have a significant effect on reading performance, as
its slope coefficient estimate had a robust 95\% posterior interval that
excluded zero. Figure 2 presents the results of the VoE analysis of the
TEXP4 effect size, based on a total of 3,600 regressions (TEXP4 effect
sizes). This figure shows that the TEXP4 effect is relatively stable and has
an overall tendency to be significant, and the larger TEXP4 effect sizes
tend to be associated with better (smaller)\ $GIC$\ statistics. The figure
also presents results of a sensitivity analysis of a hypothetical missing
confounding variable $U$, over 50 independent standard normal random samples
of $(\gamma ,\boldsymbol{\lambda })$, and shows instability of the TEXP4\
effect after accounting for this variable.

\begin{center}
------------------

Figure 2

------------------
\end{center}

\section{Conclusions}

This study introduced and illustrated regression methodology that is useful
for performing inferences of the mean dependent response. This methodology
was developed by establishing new connections between Dirichlet process
functional inference, the bootstrap, heteroscedastic-consistent sandwich
covariance estimation, ridge shrinkage regression, WLS, and VoE/sensitivity
analysis of causal effects. This study is also the first to provide
consistent sandwich covariance estimation for ridge regression. A simulation
study showed that this MDP/OLS functional methodology is competitive with
the sandwich variance estimator in terms of 95\%\ coverage rates of
posterior intervals over a large range of conditions. The former estimator
has the advantage for observed design matrices ($\mathbf{X}$) that are
multicollinear or singular. Also, the applicability of the regression
methodology was illustrated through the analysis of real data, which
involves WLS\ coefficient estimation procedures that are computationally
feasible even for very large data sets. A free software package that
implements the MDP functional methodology (menu option:\ "VoE analysis") is
available from the author's website.

Some extensions of the methods of the paper are worthy for future research.
The bootstrap approximation methodology of Section 4 yielded explicit closed
form equations for the posterior mean and covariance matrix of the OLS\
functional of the regression coefficients. This is because equations for $%
\mathbb{E}(\mathbf{n}^{\ast }\,|\,\mathbf{Z}_{n},\alpha )$ and $\mathbb{V}(%
\mathbf{n}^{\ast }\,|\,\mathbf{Z}_{n},\alpha )$ in (\ref{Boot E V}) are
available in closed form thanks to the conjugacy property of the DP prior.
This property not only allows for explicit equations for the mean and
variance of the process, but also makes it possible to correspond this mean
and variance with those (resp.)\ of the DP's P\'{o}lya urn scheme, the
latter which provides the basis for the bootstrap methodology.

In principle, the MDP\ bootstrap can be extended to other (non-conjugate)\
Bayesian nonparametric of Gibbs-type (see Leisen \& Lijoi, 2011\nocite%
{LeisenLijoi11}; Bassetti, et al. 2014\nocite{BassettiCasarinLeisen14}; Zhu
\& Leisen, 2015\nocite{ZhuLeisen15}; De Blasi et al., 2015\nocite%
{DeBlasiEtAl15}). For each of these other prior processes, the variance $%
\mathbb{V}(\mathbf{n}^{\ast }\,|\,\mathbf{Z}_{n},\alpha )$ cannot be
directly evaluated, because they do not provide explicit characterizations
of the process variance. Thus, they would require Monte Carlo simulation
methods to implement bootstrap approximations for inferences of process
functionals of interest, such as the OLS functional. Finally, the
sensitivity analysis methods of Section 5 can be extended to allow for
interactions between the treatment variable $(T,U,\underline{\boldsymbol{X}})
$, perhaps by specifying $U$ into the MDP\ baseline distribution
(VanderWeele \& Arah, 2011\nocite{VanderweeleArah11}).

\section{Acknowledgements}

The author thanks the Guest Editor Antonio Lijoi and two anonymous referees
for helpful suggestions to improve the manuscript. This research was
supported in part by NSF Grant SES-1156372. Please direct correspondence to
gkarabatsos1@gmail.com.

\bigskip 

\noindent {\Large References}

\begin{description}
\item[\noindent ] Abramowitz, M., \& Stegun, I. (1965). \textit{Handbook of
Mathematical Functions with Formulas, Graphs, and Mathematical Tables}. New
York: Dover Publications.

\item Antoniak, C. (1974). Mixtures of Dirichlet processes with applications
to Bayesian nonparametric problems. \textit{Annals of Statistics}, \textit{2}%
, 1152-1174.

\item Bassetti, F., Casarin, R., \& Leisen, F. (2014). Beta-product
dependent Pitman-Yor processes for Bayesian inference. \textit{Journal of
Econometrics}, \textit{180}, 49-72.

\item Blackwell, D., \& MacQueen, J. (1973). Ferguson distributions via P%
\'{o}lya urn schemes. \textit{Annals of Statistics}, \textit{1}, 353-355.

\item Cameron, C., \& Trivedi, P. (2005). \textit{Microeconometrics: Methods
and Applications}. New York: Cambridge University Press.

\item Chamberlain, G., \& Imbens, G. (2003). Nonparametric applications of
Bayesian inference. J\textit{ournal of Business and Economic Statistics}, 
\textit{21}, 12-18.

\item Chesher, A., \& Jewitt, I. (1987). The bias of a heteroskedasticity
consistent covariance matrix estimator. \textit{Econometrica},\textit{\ 55},
1217-1222.

\item Cifarelli, D., \& Regazzini, E. (1979). A general approach to Bayesian
analysis of nonparametric problems. The associative mean values within the
framework of the Dirichlet process. I.(Italian). \textit{Rivista di
Matematica per le Scienze Economiche e Sociali},\textit{\ 2}, 39-52.

\item Cribari-Neto, F., Ferrari, S., \& Cordeiro, G. (2000). Improved
heteroscedasticity-consistent covariance matrix estimators. \textit{%
Biometrika}, \textit{87}, 907-918.

\item DeBlasi, P., Favaro, S., Lijoi, A., Mena, R., Pr\"{u}nster, I., \&
Ruggiero, M. (2015). Are Gibbs-type priors the most natural generalization
of the Dirichlet process? \textit{IEEE Transactions on Pattern Analysis and
Machine Intelligence}, \textit{37}, 212-229.

\item DeIorio, M., M\"{u}ller, P., Rosner, G., \& MacEachern, S. (2004). An
ANOVA model for dependent random measures. \textit{Journal of the American
Statistical Association}, \textit{99}, 205-215.

\item Dunson, D., \& Park, J.-H. (2008). Kernel stick-breaking processes. 
\textit{Biometrika}, \textit{95}, 307-323.

\item Efron, B. (1979). Bootstrap methods: Another look at the jackknife. 
\textit{Annals of Statistics}, \textit{7}, 1-26.

\item Efron, B., Hastie, T., Johnstone, I., \& Tibshirani, R. (2004). Least
angle regression. \textit{Annals of Statistics},\textit{\ 32}, 407-499.

\item Eicker, F. (1963). Asymptotic normality and consistency of the least
squares estimators for families of linear regressions. \textit{Annals of
Mathematical Statistics}, \textit{34}, 447-456.

\item Eicker, F. (1967). Limit theorems for regressions with unequal and
dependent errors. In \textit{Proceedings of the Fifth Berkeley Symposium on
Mathematical Statistics and Probability} (Vol. 1, p. 59-82).

\item Fan, Y., \& Tang, C.-Y. (2013). Tuning parameter selection in high
dimensional penalized likelihood. \textit{Journal of the Royal Statistical
Society}:\textit{\ Series B}, \textit{75}, 531-552.

\item Ferguson, T. (1973). A Bayesian analysis of some nonparametric
problems. \textit{Annals of Statistics}, \textit{1}, 209-230.

\item Freedman, D. (2006). On the so-called "Huber sandwich estimator" and
"robust standard errors". \textit{American Statistician}, \textit{60},
299-302.

\item Greene, W. (2012). \textit{Econometric Analysis} (7th Ed.). Essex,
England: Pearson Education Limited.

\item Griffin, J., \& Brown, P. (2013). Some priors for sparse regression
modelling. \textit{Bayesian Analysis}, \textit{8}, 691-702.

\item Hansen, C. (2007). Asymptotic properties of a robust variance matrix
estimator for panel data when T is large. \textit{Journal of Econometrics}, 
\textit{141}, 597-620.

\item Hastie, T., Tibshiriani, R., \& Friedman, J. (2009). \textit{The
Elements of Statistical Learning: Data Mining, Inference, and Prediction}
(2nd ed.). New York: Springer-Verlag.

\item Hjort, N. (1985).\textit{\ Bayesian Nonparametric Bootstrap Confidence
Intervals }(Tech. Rep. No. 240). Department of Statistics: Stanford
University.

\item Hoerl, A., \& Kennard, R. (1970). Ridge regression: Biased estimation
for nonorthogonal problems. \textit{Technometrics}, \textit{12}, 55-67.

\item Hoff, P., \& Wakefield, J. (2013). Bayesian sandwich posteriors for
pseudo-true parameters: A discussion of "Bayesian inference with
misspecified models" by Stephen Walker. \textit{Journal of Statistical
Planning and Inference}, \textit{143}, 1638-1642.

\item Huber, P. (1967). The behavior of maximum likelihood estimates under
nonstandard conditions. In \textit{Proceedings of the Fifth Berkeley
Symposium on Mathematical Statistics and Probability}, \textit{Volume 1:
Statistics} (p. 221-233). Berkeley, CA: University of California Press.

\item Ioannidis, J. (2008). Why most discovered true associations are
inflated. \textit{Epidemiology}, \textit{19}, 640-648.

\item Ioannidis, J., Greenland, S., Hlatky, M., Khoury, M., Macleod, M.,
Moher, D., Schulz, K., \& Tibshirani, R. (2014). Increasing value and
reducing waste in research design, conduct, and analysis. \textit{The Lancet}%
, \textit{383}, 166-175.

\item James, L., Lijoi, A., \& Pr\"{u}nster, I. (2006). Conjugacy as a
distinctive feature of the Dirichlet process. \textit{Scandinavian Journal
of Statistics},\textit{\ 33}, 105-120.

\item Kim, E., Morse, A., \& Zingales, L. (2006). What has mattered to
economics since 1970. \textit{Journal of Economic Perspectives}, \textit{20}%
, 189-202.

\item Kim, J., \& Kim, J. (2012). Parametric fractional imputation for
nonignorable missing data. \textit{Journal of the Korean Statistical Society}%
, \textit{41}, 291-303.

\item Lancaster, T. (2003). \textit{A Note on Bootstraps and Robustness
(Tech. Rep.)}. Providence, RI: Brown University, Department of Economics,
No. 2006-06.

\item Leisen, F., \& Lijoi, A. (2011). Vectors of two-parameter
Poisson-Dirichlet processes. \textit{Journal of Multivariate Analysis}, 
\textit{102}, 482-495.

\item Lijoi, A., \& Pr\"{u}nster, I. (2009). Distributional properties of
means of random probability measures. \textit{Statistics Surveys}, 47-95.

\item LMT. (2012). \textit{Learning Mathematics for Teaching (LMT) Assessment%
}. Ann Arbor, MI: University of Michigan.

\item MacKinnon, J. (2013). Thirty years of heteroskedasticity-robust
inference. In X. Chen \& N. Swanson (Eds.), \textit{Recent Advances and
Future Directions in Causality, Prediction, and Specification Analysis} (p.
437-461). New York: Springer.

\item M\"{u}ller, U. (2013). Risk of Bayesian inference in misspecified
models, and the sandwich covariance matrix. \textit{Econometrica},\textit{\
81}, 1805-1849.

\item Nandram, B., \& Choi, J.-W. (2004). Nonparametric Bayesian analysis of
a proportion for a small area under nonignorable nonresponse.\textit{\
Journal of Nonparametric Statistics}, \textit{16}, 821-839.

\item Nandram, B., \& Yin, J. (2016, to appear). A nonparametric Bayesian
prediction interval for a finite population mean. \textit{Journal of
Statistical Computation and Simulation}.

\item Norets, A. (2015). Bayesian regression with nonparametric
heteroskedasticity. \textit{Journal of Econometrics}, \textit{185}, 409-419.

\item Patel, C., Burford, B., \& Ioannidis, J. (2015). Assessment of
vibration of e\textcurrency ects due to model specification can demonstrate
the instability of observational associations. \textit{Journal of Clinical
Epidemiology},\textit{\ 68}, 1046-1058.

\item Poirier, D. (2011). Bayesian interpretations of heteroskedastic
consistent covariance estimators using the informed Bayesian bootstrap. 
\textit{Econometric Reviews}, \textit{30}, 457-468.

\item Regazzini, E., Guglielmi, A., \& Nunno, G. D. (2002). Theory and
numerical analysis for exact distributions of functionals of a Dirichlet
process. \textit{Annals of Statistics}, \textit{30}, 1376-1411.

\item Rosenbaum, P. (2002). Observational Studies (2nd Ed.). New York:
Springer-Verlag.

\item Rubin, D. (1981). The Bayesian bootstrap. \textit{Annals of Statistics}%
, \textit{9}, 130-134.

\item Startz, R. (2015). \textit{Bayesian Heteroskedasticity-Robust
Regression (Tech. Rep.)}. University of California, Department of Economics.

\item Szpiro, A., Rice, K., \& Lumley, T. (2010). Model-robust regression
and a Bayesian 'sandwich' estimator. \textit{Annals of Applied Statistics}, 
\textit{4}, 2099-2113.

\item Taddy, M., Chen, C.-S., Yu, J., \& Wyle, M. (2015). Bayesian and
Empirical Bayesian Forests. \textit{In Proceedings of the 32nd International
Conference on Machine Learning (ICML-15)} (Vol. 37,p. 967-976). Lille,
France: International Machine Learning Society.

\item Vanderweele, T., \& Arah, O. (2011). Bias formulas for sensitivity
analysis of unmeasured confounding for general outcomes, treatments, and
confounders. \textit{Epidemiology}, \textit{22}, 42-52.

\item Wang, L., \& Krieger, A. (2006). Causal conclusions are most sensitive
to unobserved binary covariates. \textit{Statistics in Medicine}, \textit{25}%
, 2257-2271.

\item White, H. (1980). A heteroskedasticity-consistent covariance matrix
estimator and a direct test for heteroskedasticity. \textit{Econometrica}, 
\textit{48}, 817-838.

\item Zhu, W., \& Leisen, F. (2015). A multivariate extension of a vector of
two-parameter Poisson-Dirichlet processes. \textit{Journal of Nonparametric
Statistics}, \textit{27}, 89-105.
\end{description}

\end{document}